\documentclass[a4paper,11pt]{article}

\usepackage[height=8.85in,width=6.75in]{geometry}

\usepackage{graphicx,rotating}
\usepackage[bookmarksopen,colorlinks=true,linkcolor=light_blue,citecolor=light_pink,urlcolor=dark_red,linktoc=all]{hyperref}
\usepackage{amsmath,mathtools,amsfonts,amssymb,slashed,braket,bm,bbm,cite,xcolor,bbold,physics,xfrac}
\usepackage[footnotesize]{caption}
\usepackage{chngcntr}
\counterwithin{equation}{section}
\usepackage{comment}
\usepackage{array}
\usepackage{extarrows}

\usepackage{fourier}

\usepackage{stmaryrd}
\usepackage{trimclip}
\usepackage[framemethod=default]{mdframed}
\newmdenv[skipabove=7pt,
skipbelow=7pt,
rightline=false,
leftline=false,
topline=false,
bottomline=false,
backgroundcolor=gray!10,
linecolor=gray,
innerleftmargin=5pt,
innerrightmargin=5pt,
innertopmargin=5pt,
innerbottommargin=5pt,
leftmargin=0cm,
rightmargin=0cm,
linewidth=4pt]{eBox}
\newmdenv[skipabove=7pt,
skipbelow=7pt,
rightline=true,
leftline=true,
topline=true,
bottomline=true,
backgroundcolor=white,
linecolor=gray,
innerleftmargin=5pt,
innerrightmargin=5pt,
innertopmargin=5pt,
innerbottommargin=5pt,
leftmargin=0cm,
rightmargin=0cm,
linewidth=1pt]{eBox2}

\DeclareMathOperator{\diag}{diag}

\usepackage{multicol}
\definecolor{dark_red}{rgb}{0.7, 0., 0.}
\definecolor{light_pink}{rgb}{1,0.4,0.4}
\definecolor{light_blue}{rgb}{0.284602,0.317763,0.963947}

\newcommand{\RC}[2]{\mathbb{R}^{(#1|#2)}}
\newcommand{\tilRC}[2]{\hat{\mathbb{R}}^{(#1|#2)}}
\newcommand{\hatRC}[2]{\check{\mathbb{R}}^{(#1|#2)}}

\newcommand{\tilRCinv}[2]{\hat{\mathbb{R}}^{(#1|#2)^{-1}}}
\newcommand{\TC}[2]{\mathbb{T}^{(#1|#2)}}
\newcommand{\tilTC}[2]{\hat{\mathbb{T}}^{(#1|#2)}}
\newcommand{\hatTC}[2]{\check{\mathbb{T}}^{(#1|#2)}}
\newcommand{\TCinv}[2]{\mathbb{T}^{(#1|#2)^{-1}}}
\newcommand{\hatTCinv}[2]{\check{\mathbb{T}}^{(#1|#2)^{-1}}}
\newcommand{\tilTCinv}[2]{\hat{\mathbb{T}}^{(#1|#2)^{-1}}}

\begin{document}
\hypersetup{pageanchor=false}
\begin{titlepage}

\begin{center}

\hfill KEK-TH-2693\\
\hfill KEK-QUP-2025-0004\\
\hfill TU-1245\\
\vskip 0.8in
{\Huge \bfseries Multilayered Aspects of Casimir Energy}
\vskip .8in
{\Large
Kyohei Mukaida$^{a,b}$, Hideo Iizuka$^{c}$, Kazunori Nakayama$^{d,c}$}

\vskip .3in
\begin{tabular}{ll}
$^a$& \!\!\!\!\!\emph{Theory Center, IPNS, KEK, 1-1 Oho, Tsukuba, Ibaraki 305-0801, Japan}\\
$^b$& \!\!\!\!\!\emph{Graduate University for Advanced Studies (Sokendai), }\\[-.3em]
& \!\!\!\!\!\emph{1-1 Oho, Tsukuba, Ibaraki 305-0801, Japan}\\
$^c$& \!\!\!\!\!\emph{International Center for Quantum-field Measurement Systems}\\[-.15em]
& \!\!\!\!\!\emph{for Studies of the Universe and Particles (QUP), KEK, Tsukuba, Ibaraki 305-0801, Japan}\\
$^d$& \!\!\!\!\!\emph{Department of Physics, Tohoku University, Sendai 980-8578, Japan}\\
\end{tabular}

\end{center}

\vskip .6in

\begin{abstract}
\noindent
We give a robust formulation to calculate the Casimir energy and Casimir force for plane-parallel multilayer setups with general dielectric constants. 
We derive recursion relations for multilayer reflection and transmission coefficients in the most general setups, which are essential ingredients for evaluating the Casimir energy.  
With the use of complex analysis techniques involving the argument principle, we carefully treat and subtract UV divergences and make clear the relation between the subtraction procedure and an actual physical setup.
We also clarify the physical operational meaning of \textit{pole subtraction}, which is required to utilize the argument principle in the calculation of the Casimir energy.
Our formula is applicable to more general situations including chiral medium or Weyl semimetals.
\end{abstract}

\end{titlepage}

\tableofcontents
\thispagestyle{empty}
\renewcommand{\thepage}{\arabic{page}}
\renewcommand{\thefootnote}{$\natural$\arabic{footnote}}
\setcounter{footnote}{0}
\newpage
\hypersetup{pageanchor=true}

\renewcommand{\thefootnote}{$\sharp$\arabic{footnote}}
\setcounter{page}{1}
\setcounter{footnote}{0}

\section{Introduction}
\label{sec:intro}

Since the proposal in 1948~\cite{Casimir:1948dh}, there have been a variety of developments both in theoretical and experimental aspects of the Casimir force~\cite{Plunien:1986ca,Bordag:2001qi,Klimchitskaya:2009cw,Bordag:2009zz}. Casimir forces in equilibrium can be computed via the integration along the imaginary frequency axis~\cite{Lifshitz1956TheTO}, and those computations have revealed dependencies of Casimir forces on various materials ~\cite{PhysRevLett.105.060401,RevModPhys.88.045003,PhysRevA.95.022509} and geometries~\cite{PhysRevLett.104.160402,Ncom2013,PhysRevLett.120.040401}, with the possibility of repulsive forces through emerging materials such as Weyl semimetals~\cite{Wilson:2015wsa,Jiang:2018ivv}. 
Measurement techniques have been advancing, extending from static systems~\cite{PhysRevA.62.052109,Natphoto2011,nanophoto2021} to actively-controlled systems~\cite{Nature2021} in Casimir forces. These progresses have found directions of new quantum field search~\cite{PhysRevLett.116.221102,PhysRevD.95.123013} and possible applications in nanoscale mechanics~\cite{Natphoto2017,APL2021}.      

The Casimir force is deeply related to the quantum nature of the vacuum. In the quantum field theory, the electromagnetic field has zero point fluctuations with arbitrary frequency, which sum up to infinite vacuum energy.
That infinite vacuum energy is renormalized by the constant term in the Lagrangian, leading to (almost) zero vacuum energy.
In the presence of materials, such as two parallel metal plates, the electromagnetic waves should satisfy specific boundary conditions at the surface of the material. Due to the boundary condition, the allowed modes are discretized.
Then there appear difference between the vacuum energy inside and outside the metals. 
It is the origin of the Casimir force.
In this sense the measurement of the Casimir force directly proves the quantum vacuum.
When it comes to theoretical computation, however, it is involved due to the complication of the actual measurement setups and apparent divergence from the quantum field nature.
For an application to realistic setups, we may want a formula of the Casimir force for multilayer systems consisting of materials with general properties.
The Casimir force in multilayer system has been calculated in Ref.~\cite{PhysRevA.52.297,Klimchitskaya:2000zz,Tomas:2002swv,PhysRevLett.118.126101}. In particular, the general result for general $n$-layer system has been presented in Ref.~\cite{Tomas:2002swv} and generalized to the finite temperature case in Ref.~\cite{Raabe_2003}. 
Ref.~\cite{Tomas:2002swv} calculated the spatial component of the electromagnetic stress tensor in each layer recursively by using the Green function.

In this paper we take a different approach. 
Rather than starting from the force itself, we first calculate the Casimir energy in the multilayer system with general dielectric constants.
Since the Casimir energy stems from zero-point quantum field fluctuations, it is essential to subtract or renormalize the divergence to obtain a physical quantity. 
We will carefully subtract divergences to make use of the argument principle, which is not much taken care of in the most literature, but is directly related to the physical setup.
In addition, to apply the argument principle, we need to take care of poles frequently omitted in the calculation of the Casimir energy in the literatures, which have a clear operational meaning as we show in this paper.

Given recent increasing interests on unconventional Casimir force setups such as chiral medium, Weyl semimetals or axion electrodynamics~\cite{Wilson:2015wsa,Jiang:2018ivv,Fukushima:2019sjn,Farias:2020qqp,Oosthuyse:2023mbs,Favitta:2023hlx,Ema:2023kvw}, we aim to provide a general formalism that is applicable to these cases without ambiguity on the relation between theoretical calculation and an actual setup.

This paper is organized as follows.
In Sec.~\ref{sec:setup}, as a preparation to the following sections, we summarize our general setup and give formulae for recursion relation for reflection/transmission coefficients. 
In Sec.~\ref{sec:cas_energy_mul} we calculate the Casimir energy by making use of the argument principle. 
In doing so, we carefully treat the divergence and give clear physical interpretation of the subtraction procedure.
We also calculate the Casimir force as a trivial application of our formula. 
In Sec.~\ref{sec:phys_int} we give an operational meaning of our formula.
Sec.~\ref{sec:conc} is devoted conclusions.

\section{Preliminary}
\label{sec:setup}

\begin{figure}[t]
    \centering
    \includegraphics[width=0.5\linewidth]{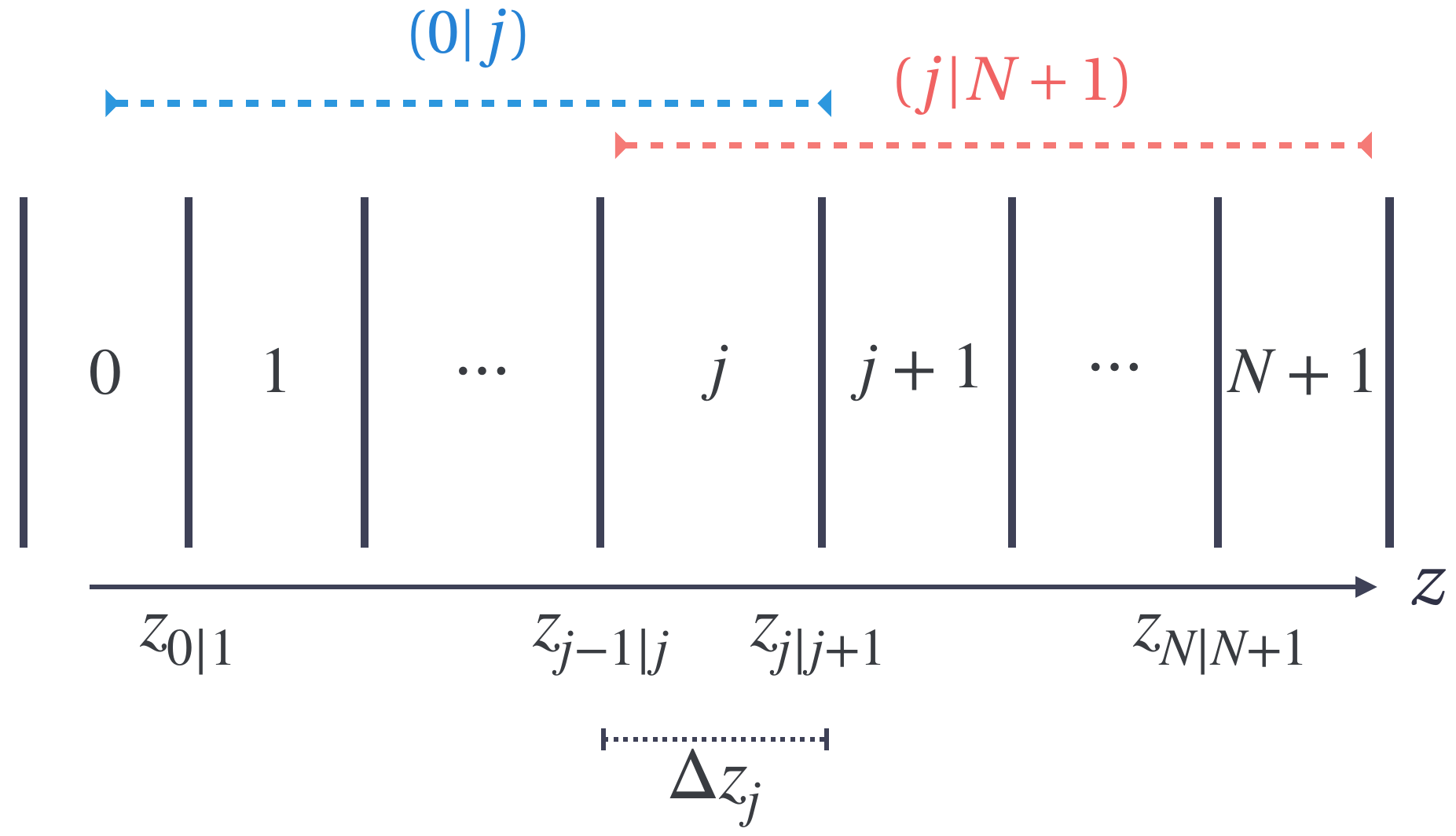}
    \caption{Schematic figure of a multilayer consisting of $(N+2)$ regions with $(N+1)$ interfaces discussed in this paper.}
    \label{fig:multilayer}
\end{figure}

\subsection{Setup}

In this paper, we study a system of $(N+2)$ regions with $(N+1)$ interfaces each of which is an infinite plane  located at $z = z_{i|i+1}$ for $i = 0, \cdots, N$ (see Fig.~\ref{fig:multilayer}).
The width of each layer $i$ is given by $\Delta z_i \equiv z_{i|i+1} - z_{i-1|i}$ for $i = 1, \cdots, N$. 
The leftmost (rightmost) region for $z < z_{0|1}$ ($z > z_{N|N+1}$) serves as a boundary of the system.
Each interface is assumed to be a movable wall.
We denote two neighboring layers separated by an interface of $z_{j|j+1}$ as $(j|j+1)$.
Note that $(j|j+1)$ does not involve its boundaries at $z_{j-1|j}$ and $z_{j+1|j+2}$.
We extend this notation to include the stack of layers from $i$ to $j$ as $(i|j)$.

Suppose that a field $\phi (t, \bm{x})$ obeys the following equation of motion in each region $j$:
\begin{equation} \label{eq:phi_eom}
    0 = \qty[ \qty( \omega^2 - \bm{k}^2 ) \delta_{\lambda\lambda'} + i \Pi^{(j)}_{\lambda \lambda'} (\omega, \bm{k}) ] \phi_{\lambda'} (\omega, \bm{k}).
\end{equation}
Here the subscript $\lambda$ represents a possible polarization of $\phi$ that runs through $\lambda = 1, \cdots, n$ with $n$ being a degree of freedom.
For simplicity, we assume that the each layer $j$ is uniform and has translational/rotational invariance in the $x$-$y$ plane.
We, however, allow a vectorial dependence in the $z$ direction, $\vec b$, for instance, a Weyl semimetal with a vector that connects two Weyl nodes in the Brillouin zone pointing to the $z$ direction.
The in-medium correction of the region $j$ is denoted by $\Pi^{(j)}_{\lambda \lambda'}$, which is generally a matrix of the polarization.
As we are interested in a system which consists of the stack of layers along the $z$-axis, it is convenient to express the solution of Eq.~\eqref{eq:phi_eom} as a function of its energy, $\omega$, and momenta perpendicular to the $z$-axis, $\bm{k}_\perp$, namely
\begin{equation}
    \text{transverse momentum:} ~~ k_{z \lambda}^{(j)} (\omega, \bm{k}_\perp), \qquad
    \text{polarization:} ~~ \epsilon_{R\lambda}^{(j)} (\omega, \bm{k}_\perp), \quad \epsilon_{L\lambda}^{(j)} (\omega, \bm{k}_\perp),
\end{equation}
where the transverse momentum $k_{z \lambda}^{(j)}$ is taken to be positive, and the subscripts $R$ and $L$ indicate the right- and left-moving modes respectively.
The presence of the vector $\vec{b}$ in the $z$ direction can be incorporated in the transverse momentum $k_{z \lambda}^{(j)}$, which splits two modes that are otherwise degenerate by the parity symmetry, \textit{i.e.,} each mode $k^{(j)}_{\bullet +}$ is shifted to $k^{(j)}_{\bullet -}$ by flipping the sign of $\vec{b}$.
A primary example is the Weyl semimetal, where the sign of each mode is determined by inner product of $\vec{b}$ and the helicity as $\vec{b}$ transforms as a vector while the helicity transforms as an axialvector.

\begin{figure}[t]
    \centering
    {\includegraphics[width=0.35\linewidth]{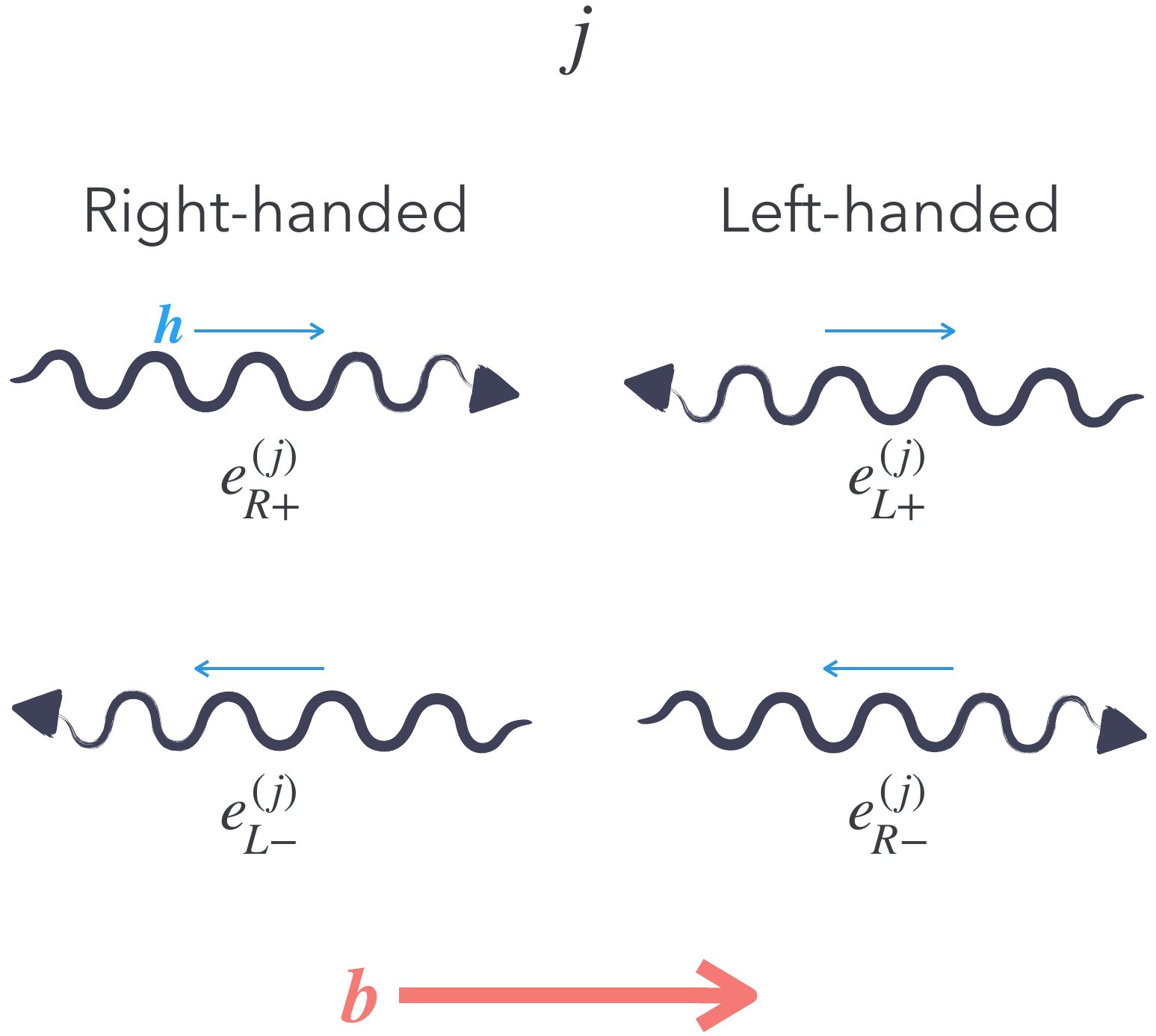}} \qquad
    {\includegraphics[width=0.35\linewidth]{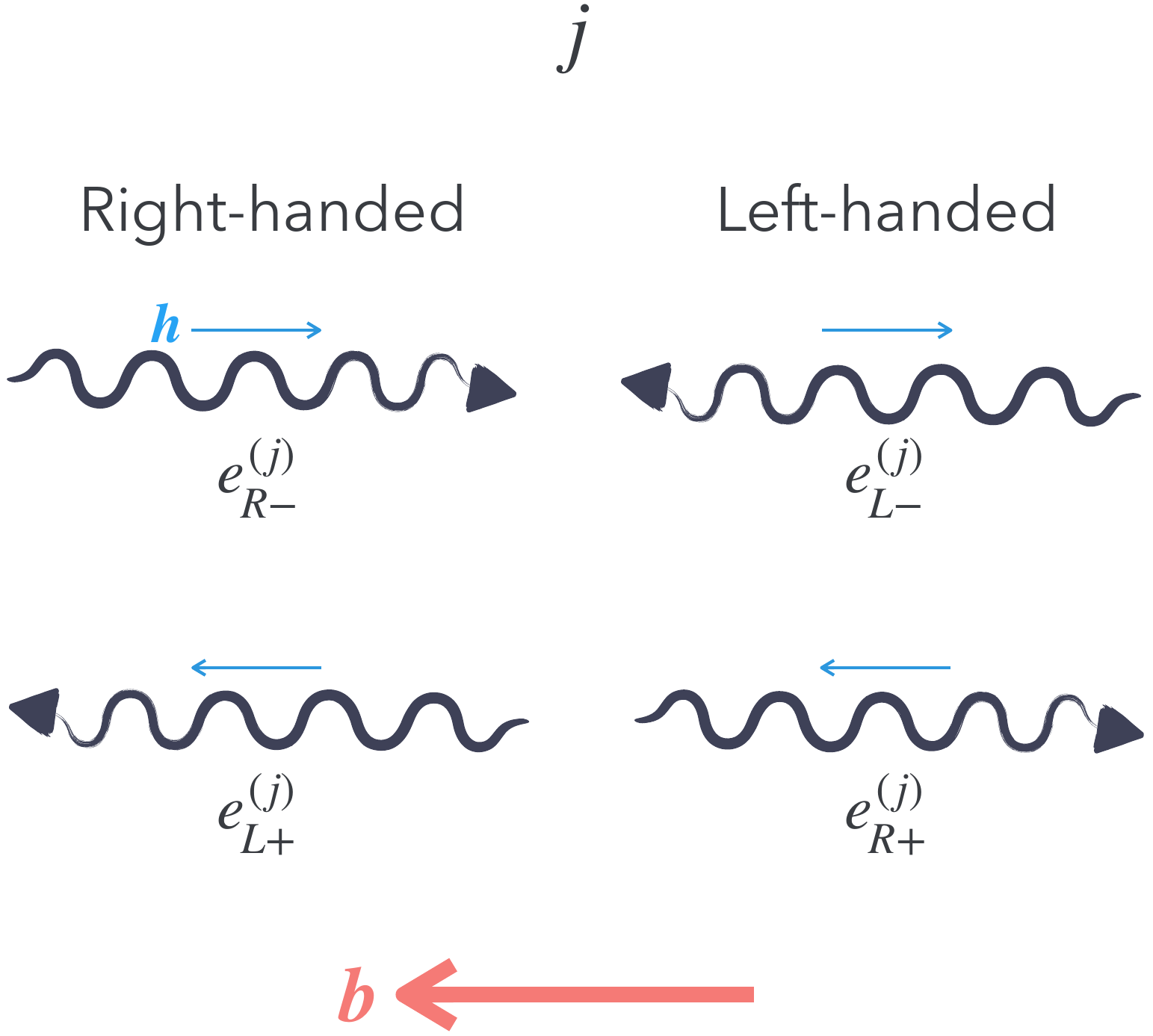}}
    \caption{
        \textbf{(Left)} The right-handed and left-handed modes by means of the polarization vector $e_{\bullet \pm}$ with respect to the right- and left-moving modes.
        The $\pm$ sign corresponds to the sign of the inner product of $\bm{b}$ and the helicity $\bm{h}$.
        \textbf{(Right)} The same figure with an opposite sign of $\bm{b}$.
    }
    \label{fig:left_right}
\end{figure}

The field $\phi$ can be expanded as
\begin{equation}
    \phi ( t, \bm{x}_\perp, z) = 
    \int_{\omega, \bm{k}_\perp} \!\!\!\!\! \!\!\!\!\! \!\!\! e^{- i (\omega t + \bm{k}_\perp \cdot \bm{x}_\perp)} 
    \sum_\lambda 
    \qty[ 
        \phi_{R\lambda}^{(j)} (\omega, \bm{k}_\perp) \epsilon_{R \lambda}^{(j)} (\omega, \bm{k}_\perp) e^{i k_{z\lambda}^{(j)} (\omega, \bm{k}_\perp) z} 
        +
        \phi_{L\lambda}^{(j)} (\omega, \bm{k}_\perp) \epsilon_{L \lambda}^{(j)} (\omega, \bm{k}_\perp) e^{- i k_{z\lambda}^{(j)} (\omega, \bm{k}_\perp) z} 
    ],
\end{equation}
for each region $j$, \textit{i.e.,} $z_{j-1|j} \leqslant z \leqslant z_{j|j+1}$.
The wave functions $\phi_{R\lambda}^{(j)}$ and $\phi_{L\lambda}^{(j)}$ should be connected so that they fulfill the reflection and transmission properties at each interface, which will be discussed in the next subsection.
The polarization basis can be taken arbitrary if $k_{z\lambda}$ is independent of the polarization $\lambda$. This is not the case for Weyl semimetal with $\vec b$, for example.
Then one should note that this mode expansion defines the polarization $\lambda$ for the right- and left-moving modes by the same transverse momentum $k_{z\lambda}^{(j)}$.
This implies that the handedness of the right- and left-moving modes for the same $\lambda$ is opposite to each other, \textit{i.e.,} $e_{R\bullet +}$ and $e_{L\bullet -}$ are the right-handed, while $e_{R\bullet -}$ and $e_{L\bullet +}$ are the left-handed (see Fig.~\ref{fig:left_right}).

\subsection{Recursion relation for reflection/transmission coefficients}

\begin{figure}[t]
    \centering
    \includegraphics[width=0.47\linewidth]{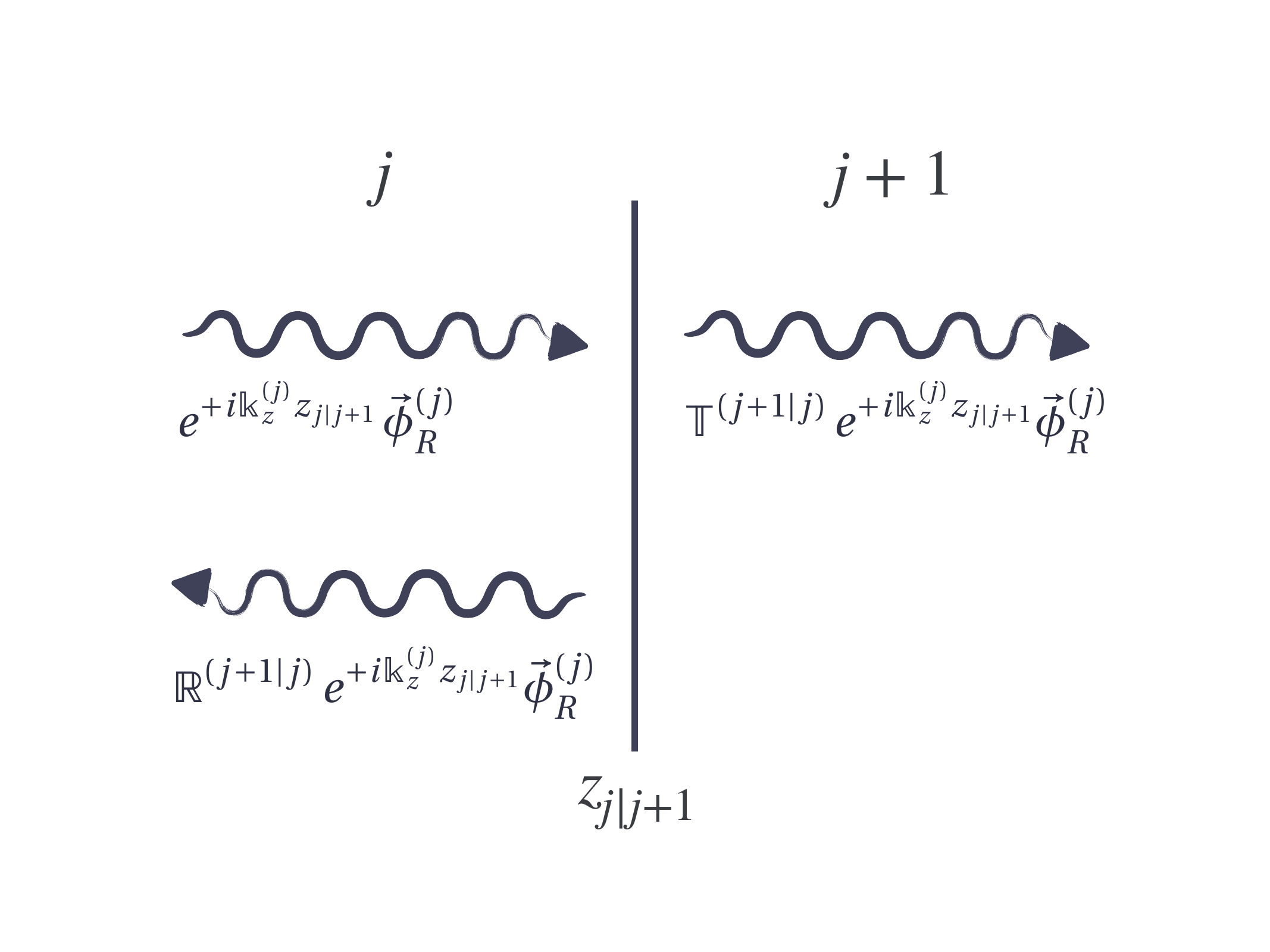}
    \includegraphics[width=0.47\linewidth]{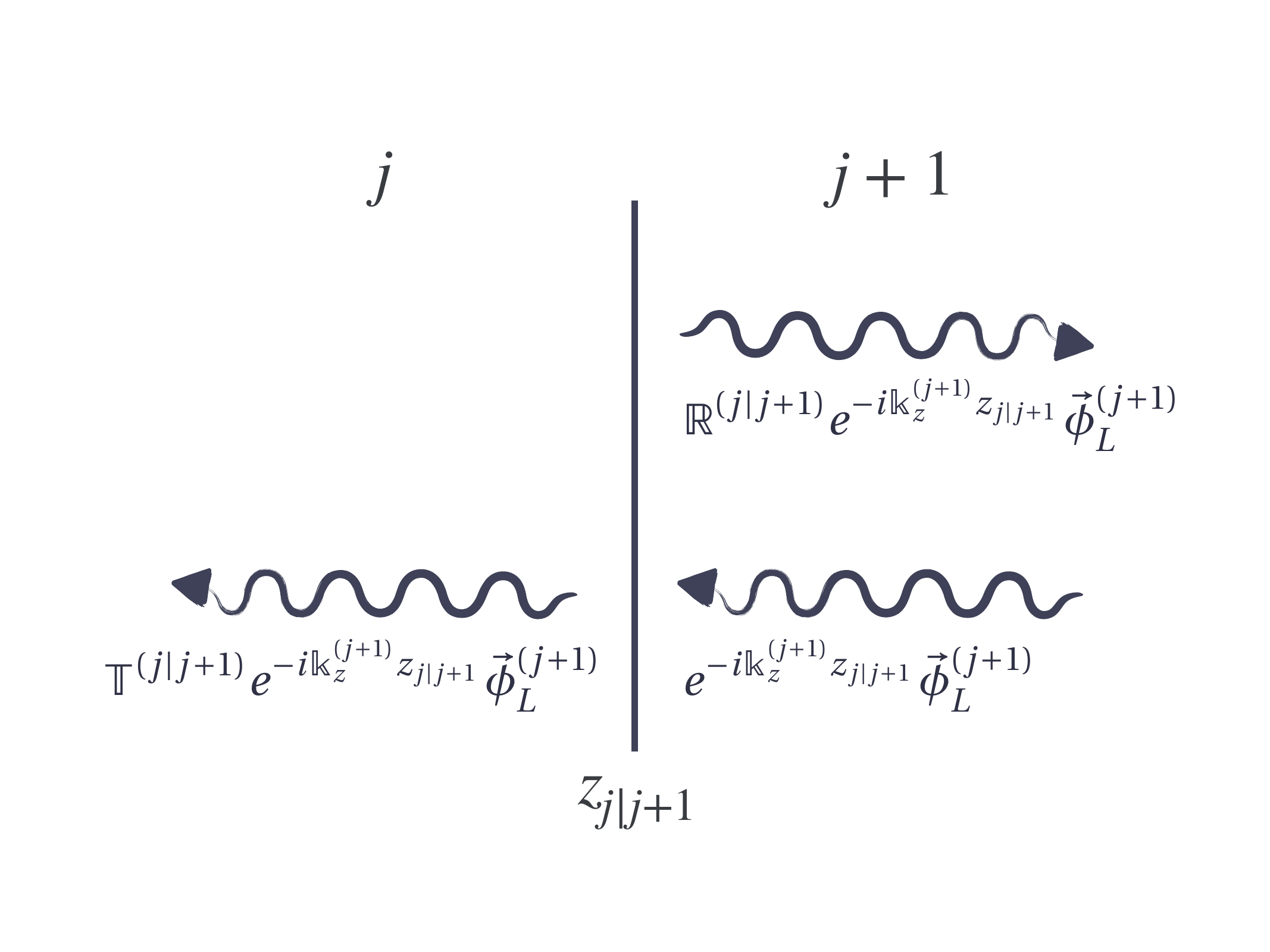}
    \caption{
        \textbf{(Left)} Injection of a wave from the $j$-layer to the $(j+1)$-layer.
        \textbf{(Right)} Injection of a wave from the $(j+1)$-layer to the $j$-layer.
    }
    \label{fig:j_j+1}
\end{figure}

Let us first discuss the connection of two solutions in the $j$-region and $(j+1)$-region at the interface $z = z_{j|j+1}$. 
The transfer matrix relates the wave functions in the $j$- and $(j+1)$-regions as follows
\begin{equation}
    \begin{pmatrix}
        \vec{\phi}_{R}^{(j+1)}
        \\
        \vec{\phi}_{L}^{(j+1)}
    \end{pmatrix}
    = 
    \mathbb{M}_{j+1|j}
    \begin{pmatrix}
        \vec{\phi}_{R}^{(j)}
        \\
        \vec{\phi}_{L}^{(j)}
    \end{pmatrix},
\end{equation}
where we collectively denote the wave function of each polarization as  $\vec{\phi}_\bullet^{(j)} = (\phi_{\bullet 1}^{(j)}, \cdots, \phi_{\bullet n}^{(j)})^\top$.
Note that the transfer matrix $\mathbb{M}_{j+1|j}$ is defined with respect to $\lambda$ not by helicity.
Suppose that we inject a wave of $e^{+i \mathbb{k}_{z}^{(j)} z}\, \vec{\phi}_{R}^{(j)}$ from the region $j$ into the interface of $z = z_{j|j+1}$ (see also Fig.~\ref{fig:j_j+1}). 
Here we introduce the following shorthanded notation for brevity: $\mathbb{k}_z^{(j)} \equiv \diag( k_{z1}^{(j)}, \cdots, k_{zn}^{(j)} )$.
By means of the reflection matrix, $\RC{j+1}{j}$, and transmission matrix, $\TC{j+1}{j}$, the reflected and transmitted waves can be expressed as 
$e^{- i \mathbb{k}_{z}^{(j)} z_{j|j+1}} \vec{\phi}_L^{(j)}= 
\RC{j+1}{j}\, e^{+i \mathbb{k}_{z}^{(j)} z_{j|j+1}} \vec{\phi}_{R}^{(j)}$ and
$e^{+ i \mathbb{k}_{z}^{(j+1)} z_{j|j+1}}\,\vec{\phi}_R^{(j+1)}
=
\mathbb{T}^{(j+1|j)}\,e^{+i \mathbb{k}_{z}^{(j)} z_{j|j+1}}
\vec{\phi}_{R}^{(j)}$, namely
\begin{equation}\label{eq:jtoj+1}
    \begin{pmatrix}
        e^{- i \mathbb{k}_{z}^{(j+1)} z_{j|j+1}}\, 
        \mathbb{T}^{(j+1|j)}
        e^{+i \mathbb{k}_{z}^{(j)} z_{j|j+1}} \,
        \vec{\phi}_{R}^{(j)} \\
        0
    \end{pmatrix}
    = 
    \mathbb{M}_{j+1|j}
    \begin{pmatrix}
        \vec{\phi}_{R}^{(j)}\\ 
        e^{+ i \mathbb{k}_{z}^{(j)} z_{j|j+1}}
        \, \RC{j+1}{j} 
        e^{+i \mathbb{k}_{z}^{(j)} z_{j|j+1}} \,
        \vec{\phi}_{R}^{(j)} 
    \end{pmatrix}.
\end{equation}
The concrete forms are obtained once we specify the properties of the region $j$ and $j+1$ explicitly, 
On the other hand, if we inject a wave of $e^{- i \mathbb{k}_z^{(j+1)}z}\, \vec{\phi}_L^{(j+1)}$ from the region $j+1$ into the interface $z = z_{j|j+1}$, the reflected and transmitted waves can be expressed as
\begin{equation}\label{eq:j+1toj}
    \mathbb{M}_{j|j+1}
    \begin{pmatrix}
        e^{- i \mathbb{k}_{z}^{(j+1)} z_{j|j+1}}
        \, \RC{j}{j+1}
        e^{-i \mathbb{k}_{z}^{(j+1)} z_{j|j+1}} \,
        \vec{\phi}_{L}^{(j+1)} 
        \\ 
        \vec{\phi}_{L}^{(j+1)}
    \end{pmatrix}
    =
    \begin{pmatrix}
        0 \\ 
        e^{+i \mathbb{k}_{z}^{(j)} z_{j|j+1}}
        \, \TC{j}{j+1}
        e^{-i \mathbb{k}_{z}^{(j+1)} z_{j|j+1}} \,
        \vec{\phi}_{L}^{(j+1)}
    \end{pmatrix}.
\end{equation}
By definition, we have $\mathbb{M}_{j|j+1} \mathbb{M}_{j+1|j} = \mathbb{1}$, \textit{i.e.}, $\mathbb{M}_{j|j+1} = \mathbb{M}_{j+1|j}^{-1}$.

These relations given in Eqs.~\eqref{eq:jtoj+1} and \eqref{eq:j+1toj} are naturally extended to the reflection and transmission for a stack of layers.
Let us consider a stack of layers from $i$ to $j$ with $i < j$ denoted by $(i|j)$, and inject a wave from the region $i$ to the interface of $z_{i|j} \equiv \{z ~|~ z_{i|i+1} \leqslant z \leqslant z_{j-1|j} \}$.
By introducing the effective reflection and transmission matrices for the stack of layers $(i|j)$, we can write down the reflected wave function in the region $i$ and the transmitted wave function in the region $j$ as follows
\begin{equation}\label{eq:itoj}
    \begin{pmatrix}
        e^{-i \mathbb{k}_{z}^{(j)} z_{j-1|j}} 
        \, \mathbb{T}^{(j|i)}
        e^{+i \mathbb{k}_{z}^{(i)} z_{i|i+1}} \,
        \vec{\phi}_{R}^{(i)} \\
        0
    \end{pmatrix}
    = 
    \mathbb{M}_{j|i}
    \begin{pmatrix}
        \vec{\phi}_{R}^{(i)}\\ 
        e^{+i \mathbb{k}_{z}^{(i)} z_{i|i+1}}
        \, \RC{j}{i}
        e^{+i \mathbb{k}_{z}^{(i)} z_{i|i+1}} \,
        \vec{\phi}_{R}^{(i)} 
    \end{pmatrix},
\end{equation}
where $\mathbb{M}_{j|i} \equiv \mathbb{M}_{j|j-1} \mathbb{M}_{j-1|j-2} \cdots \mathbb{M}_{i+1|i}$.
For the opposite process, \textit{i.e.,} incoming wave from the region $j$, we have a similar expression
\begin{equation}\label{eq:jtoi}
    \mathbb{M}_{i|j}
    \begin{pmatrix}
        e^{-i \mathbb{k}_{z}^{(j)} z_{j-1|j}}
        \, \RC{i}{j} 
        e^{-i \mathbb{k}_{z}^{(j)} z_{j-1|j}} \,
        \vec{\phi}_{L}^{(j)} \\ 
        \vec{\phi}_{L}^{(j)}
    \end{pmatrix}
    =
    \begin{pmatrix}
        0\\ 
        e^{+i \mathbb{k}_{z}^{(i)} z_{i|i+1}}
        \, \mathbb{T}^{(i|j)}
        e^{-i \mathbb{k}_{z}^{(j)} z_{j-1|j}} \,
        \vec{\phi}_{L}^{(j)} 
    \end{pmatrix}.
\end{equation}

From Eqs.~(\ref{eq:itoj}) and (\ref{eq:jtoi}), we obtain
\begin{equation}
	 \mathbb{M}_{i|j} =
	  \begin{pmatrix}
         e^{-i \mathbb{k}_{z}^{(i)} z_{i|i+1}}\,\mathbb{T}^{(j|i)^{-1}}\,e^{+i \mathbb{k}_{z}^{(j)} z_{j-1|j}}  & 
        -e^{-i \mathbb{k}_{z}^{(i)} z_{i|i+1}}\,\mathbb{T}^{(j|i)^{-1}}\, \RC{i}{j}\, e^{-i \mathbb{k}_{z}^{(j)} z_{j-1|j}} \\ 
        e^{+i \mathbb{k}_{z}^{(i)} z_{i|i+1}}\,\RC{j}{i}\, \mathbb{T}^{(j|i)^{-1}}\, e^{+i \mathbb{k}_{z}^{(j)} z_{j-1|j}}   &
        e^{+i \mathbb{k}_{z}^{(i)} z_{i|i+1}}\,\left[\mathbb{T}^{(i|j)} - \RC{j}{i} \mathbb{T}^{(j|i)^{-1}} \RC{i}{j}  \right]\, e^{-i \mathbb{k}_{z}^{(j)} z_{j-1|j}}  
    \end{pmatrix}.
    \label{Mij}
\end{equation}
and
\begin{align}
	 \mathbb{M}_{j|i} =
	  \begin{pmatrix}
        e^{-i \mathbb{k}_{z}^{(j)} z_{j-1|j}}\,\left[\mathbb{T}^{(j|i)} - \RC{i}{j} \mathbb{T}^{(i|j)^{-1}} \RC{j}{i} \right]\, e^{+i \mathbb{k}_{z}^{(i)} z_{i|i+1}}  & 
        e^{-i \mathbb{k}_{z}^{(j)} z_{j-1|j}}\,\RC{i}{j}\, \mathbb{T}^{(i|j)^{-1}}\,e^{-i \mathbb{k}_{z}^{(i)} z_{i|i+1}} \\ 
        -e^{+i \mathbb{k}_{z}^{(j)} z_{j-1|j}}\,
        \mathbb{T}^{(i|j)^{-1}}\, \RC{j}{i}\,e^{+i \mathbb{k}_{z}^{(i)} z_{i|i+1}}  &
         e^{+i \mathbb{k}_{z}^{(j)} z_{j-1|j}}\,\mathbb{T}^{(i|j)^{-1}}\,e^{-i \mathbb{k}_{z}^{(i)} z_{i|i+1}} 
    \end{pmatrix}.
    \label{Mji}
\end{align}
for $i < j$. We can explicitly check that $ \mathbb{M}_{i|j}  \mathbb{M}_{j|i} = \mathbb{1}$. Note that the expression (\ref{Mji}) can also be obtained from (\ref{Mij}) with the following replacement: $i\leftrightarrow j$, $\mathbb{k}_{z} \leftrightarrow -\mathbb{k}_{z}$ and $L \leftrightarrow R$.

As will be shown shortly, the effective reflection and transmission matrices fulfill the recursion relations that follow from the definition.
Pick up an intermediate layer $k$ with $i < k < j$ and consider a wave injected from the layer $j$.
Dividing the transfer matrix into $\mathbb{M}_{i|j} = \mathbb{M}_{i|k} \mathbb{M}_{k|j}$,
we multiply $\mathbb{M}_{i|k}^{-1} = \mathbb{M}_{k|i}$ to the both-hand sides. 
The right-hand side of Eq.~\eqref{eq:jtoi} becomes
\begin{equation}
    \mathbb{M}_{k|i}
    \begin{pmatrix}
        0 \\ 
       \vec{\tilde{\phi}}_L^{(j)}
    \end{pmatrix}
    =
    \begin{pmatrix}
        e^{-i \mathbb{k}_{z}^{(k)} z_{k-1|k}}
        \, \RC{i}{k}
        \mathbb{T}^{(i|k)^{-1}} 
        e^{-i \mathbb{k}_{z}^{(i)} z_{i|i+1}}\,
        \vec{\tilde\phi}_{L}^{(j)} \\ 
        e^{+i \mathbb{k}_{z}^{(k)} z_{k-1|k}}
        \,\mathbb{T}^{(i|k)^{-1}}
        e^{-i \mathbb{k}_{z}^{(i)} z_{i|i+1}} \,
        \vec{\tilde\phi}_{L}^{(j)} 
    \end{pmatrix}
\end{equation}
with $\vec{\tilde{\phi}}_L^{(j)} \equiv e^{+i \mathbb{k}_{z}^{(i)} z_{i|i+1}}\,\mathbb{T}^{(i|j)}e^{-i \mathbb{k}_{z}^{(j)} z_{j-1|j}}\,\vec{\phi}_{L}^{(j)}$.
Then, we further multiply $\mathbb{M}_{k|j}^{-1} = \mathbb{M}_{j|k}$, which leads to
\begin{align} \label{eq:ktoj}
    &
    \mathbb{M}_{j|k} 
    \begin{pmatrix}
        e^{-i \mathbb{k}_{z}^{(k)} z_{k-1|k}}
        \, \RC{i}{k}
        \mathbb{T}^{(i|k)^{-1}} 
        e^{-i \mathbb{k}_{z}^{(i)} z_{i|i+1}}\,
        \vec{\tilde\phi}_{L}^{(j)} \\ 
        e^{+i \mathbb{k}_{z}^{(k)} z_{k-1|k}}
        \,\mathbb{T}^{(i|k)^{-1}}
        e^{-i \mathbb{k}_{z}^{(i)} z_{i|i+1}} \,
        \vec{\tilde\phi}_{L}^{(j)} 
    \end{pmatrix}\\ \nonumber
    &=
    \begin{pmatrix}
        e^{-i \mathbb{k}_{z}^{(j)} z_{j-1|j}}
        \qty{
            \RC{k}{j}
            \mathbb{T}^{(k|j)^{-1}}
            \qty[
            e^{-i \mathbb{k}_{z}^{(k)} \Delta z_{k}}
            -
            \RC{j}{k}
            e^{i \mathbb{k}_z^{(k)}\Delta z_k}
            \RC{i}{k}
            ]
            +
            \mathbb{T}^{(j|k)}
            e^{i \mathbb{k}_{z}^{(k)} \Delta z_{k}} \, 
            \RC{i}{k}
            }
            \mathbb{T}^{(i|k)^{-1}}\,
            e^{-i \mathbb{k}_{z}^{(i)} z_{i|i+1}}\,
        \vec{\tilde\phi}_{L}^{(j)} 
        \\
        e^{+i \mathbb{k}_{z}^{(j)} z_{j-1|j}}
        \mathbb{T}^{(k|j)^{-1}}
        \qty[
            e^{- i \mathbb{k}_z^{(k)}\Delta z_k}
            -
            \RC{j}{k}
            e^{i \mathbb{k}_z^{(k)}\Delta z_k}
            \RC{i}{k}
        ]
        \mathbb{T}^{(i|k)^{-1}}
        e^{-i \mathbb{k}_{z}^{(i)} z_{i|i+1}}
        \vec{\tilde\phi}_{L}^{(j)}
    \end{pmatrix}.
\end{align}
On the other hand, the left-hand side of Eq.~\eqref{eq:jtoi} now becomes
$(
        e^{-i \mathbb{k}_{z}^{(j)} z_{j-1|j}}
        \, \RC{i}{j} 
        e^{-i \mathbb{k}_{z}^{(j)} z_{j-1|j}} \,
        \vec{\phi}_{L}^{(j)}, 
        \vec{\phi}_{L}^{(j)})^\top
$.
By comparing this to Eq.~\eqref{eq:ktoj}, we obtain the recursion relations
\begin{eBox}
\begin{align} \label{eq:recur_T}
    \mathbb{T}^{(i|j)} 
    &= 
    \mathbb{T}^{(i|k)}
        \qty[
            e^{- i \mathbb{k}_z^{(k)}\Delta z_k}
            -
            \RC{j}{k}
            e^{i \mathbb{k}_z^{(k)}\Delta z_k}
            \RC{i}{k}
        ]^{-1}
    \mathbb{T}^{(k|j)}, \\ \label{eq:recur_R}
    \RC{i}{j}
    &=
    \RC{k}{j} + 
    \mathbb{T}^{(j|k)}
    e^{i \mathbb{k}_{z}^{(k)} \Delta z_{k}} \, 
    \RC{i}{k}
        \mathbb{T}^{(i|k)^{-1}}
        \mathbb{T}^{(i|j)}.
\end{align}
\end{eBox}
Although we have assumed $i < j$ in the derivation of Eqs.~\eqref{eq:recur_T} and \eqref{eq:recur_R}, the same recursion relations hold even for $i > j$.
These recursion equations allow us to express the effective reflection and transmission matrices of the individual interface, $\RC{j+1}{j}$, $\RC{j}{j+1}$, $\mathbb{T}^{(j+1|j)}$, and $\mathbb{T}^{(j|j+1)}$.

By construction, the left-hand side of recursion relations \eqref{eq:recur_T} and \eqref{eq:recur_R} should not depend on the choice of the layer $k$ between $i$ and $j$.
We briefly sketch the proof of this property for later use.
Let us pick up a layer $l$ located between $i$ and $k$, namely, $i < l < k < j$.
By applying the recursion relation \eqref{eq:recur_T} to $\mathbb{T}^{(i|k)}$ for the layer $l$, we obtain
\begin{align}\label{eq:T_ikj}
    \mathbb{T}^{(i|k|j)^{-1}}
    &= 
    \mathbb{T}^{(k|j)^{-1}} 
        \qty[
            e^{- i \mathbb{k}_z^{(k)}\Delta z_k}
            -
            \RC{j}{k}
            e^{i \mathbb{k}_z^{(k)}\Delta z_k}
            \RC{i}{k}
        ]
    \mathbb{T}^{(i|k)^{-1}}\\
    &=
    \mathbb{T}^{(k|j)^{-1}} 
    \qty[ e^{- i \mathbb{k}_z^{(k)}\Delta z_k}
    -
    \RC{j}{k}
    e^{i \mathbb{k}_z^{(k)}\Delta z_k}
    \RC{i}{k} ]
    \mathbb{T}^{(l|k)^{-1}}
    \qty[ e^{- i \mathbb{k}_z^{(l)}\Delta z_l}
    -
    \RC{k}{l}
    e^{i \mathbb{k}_z^{(l)}\Delta z_l}
    \RC{i}{l} ]
    \mathbb{T}^{(i|l)^{-1}}.
\end{align}
Here we write down a fake dependence of the layer $k$ in the left-hand side of Eq.~\eqref{eq:T_ikj} to make it explicit that the expression of Eq.~\eqref{eq:T_ikj} is specific to the recursion relation for the layer $k$ inserted.
On the other hand, one may first apply the recursion relation to $\mathbb{T}^{(i|j)}$ for the layer $l$, and then to $\mathbb{T}^{(l|j)}$ for the layer $k$, which gives
\begin{align}
    \mathbb{T}^{(i|l|j)^{-1}}
    &=
    \mathbb{T}^{(l|j)^{-1}}
        \qty[
            e^{- i \mathbb{k}_z^{(l)}\Delta z_l}
            -
            \RC{j}{l}
            e^{i \mathbb{k}_z^{(l)}\Delta z_l}
            \RC{i}{l}
        ]
    \mathbb{T}^{(i|l)^{-1}} \\
    &=
    \mathbb{T}^{(k|j)^{-1}}
        \qty[
            e^{- i \mathbb{k}_z^{(k)}\Delta z_k}
            -
            \RC{j}{k}
            e^{i \mathbb{k}_z^{(k)}\Delta z_k}
            \RC{l}{k}
        ]
    \mathbb{T}^{(l|k)^{-1}}
        \qty[
            e^{- i \mathbb{k}_z^{(l)}\Delta z_l}
            -
            \RC{j}{l}
            e^{i \mathbb{k}_z^{(l)}\Delta z_l}
            \RC{i}{l}
        ]
    \mathbb{T}^{(i|l)^{-1}}.
\end{align}
Again, we write down the fake dependence of the inserted layer $l$.
Hence, the required property of $\mathbb{T}^{(i|k|j)^{-1}} = \mathbb{T}^{(i|l|j)^{-1}} = \mathbb{T}^{(i|j)^{-1}}$ follows if the following equality is fulfilled 
\begin{align}
    &\qty[ e^{- i \mathbb{k}_z^{(k)}\Delta z_k}
    -
    \RC{j}{k}
    e^{i \mathbb{k}_z^{(k)}\Delta z_k}
    \RC{i}{k} ]
    \mathbb{T}^{(l|k)^{-1}}
    \qty[ e^{- i \mathbb{k}_z^{(l)}\Delta z_l}
    -
    \RC{k}{l}
    e^{i \mathbb{k}_z^{(l)}\Delta z_l}
    \RC{i}{l} ] \nonumber \\
    &=
    \qty[
            e^{- i \mathbb{k}_z^{(k)}\Delta z_k}
            -
            \RC{j}{k}
            e^{i \mathbb{k}_z^{(k)}\Delta z_k}
            \RC{l}{k}
        ]
    \mathbb{T}^{(l|k)^{-1}}
        \qty[
            e^{- i \mathbb{k}_z^{(l)}\Delta z_l}
            -
            \RC{j}{l}
            e^{i \mathbb{k}_z^{(l)}\Delta z_l}
            \RC{i}{l}
        ],
        \label{eq:consistency_rel}
\end{align}
which can be shown straightforwardly as done in Appendix.
In a similar way, one may also show the $k$-independence of the left-hand side of Eq.~\eqref{eq:recur_R}.

\begin{figure}[t]
    \centering
    \includegraphics[width=0.47\linewidth]{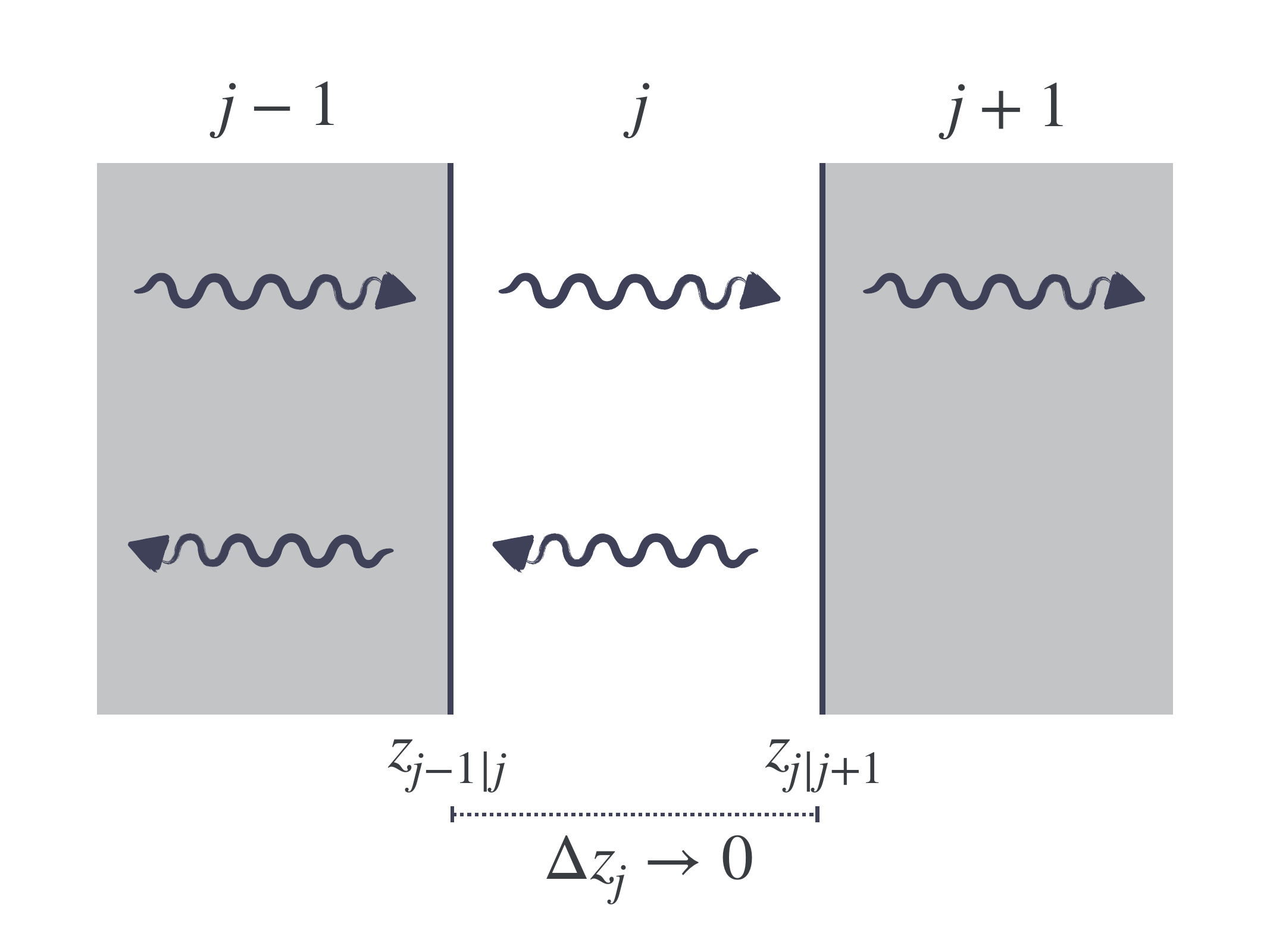}  \qquad
    \includegraphics[width=0.47\linewidth]{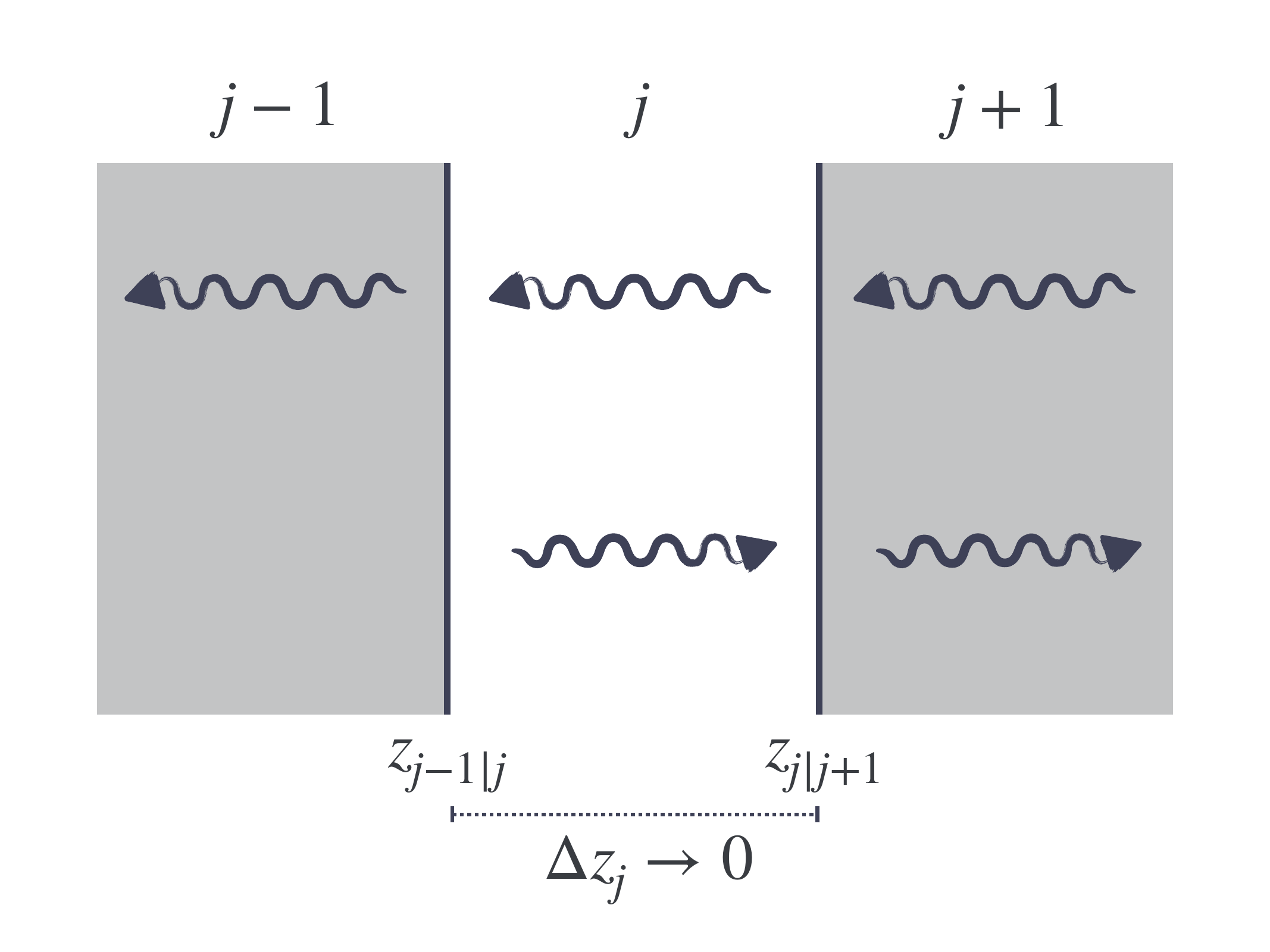}
    \vspace{-.5em}
    \caption{
        \textbf{(Left)} The artificial setup to derive the consistency conditions for the reflection and transmission coefficients in Eq.~\eqref{eq:rel_jp1toj}. We take the same medium as the layer $j+1$ for $j-1$, and merge the interface $z = z_{j-1|j}$ into $z_{j|j+1}$.
        \textbf{(Right)} The same setup as the left panel, but for Eq.~\eqref{eq:rel_jm1toj}.
    }
    \label{fig:cntrct}
\end{figure}

Before closing this section, we discuss a ``contractibility'' of interfaces assumed throughout this paper.
Let us focus on the interface placed at $z = z_{j|j+1}$.
To derive the consistency condition, we also place the same medium as a layer $j+1$ at $j-1$ and the interface $z = z_{j-1|j}$ (see the left panel of Fig.~\ref{fig:cntrct}).
We assume that the layer $j$ can be squeezed smoothly without causing any singularity by merging $z_{j-1|j} \to z_{j|j+1}$.
By considering this squeezing process, we find that the above requirement poses non-trivial relations among reflection and transmission coefficients as follows:
\begin{align}
    \mathbb{1} &= \TC{j}{j-1} \TC{j+1}{j}  + \RC{j-1}{j} \RC{j+1}{j}, \\
    0 &= \TC{j-1}{j} \RC{j+1}{j} + \RC{j}{j-1} \TC{j+1}{j}.
\end{align}
By construction, the property of the interface $z_{j-1|j}$ should be almost the same as that of $z_{j|j+1}$ except for the direction of $\vec{b}$.
Hence, for the same $\lambda$, the helicity of the incoming wave from $j+1$ to $j$ and that from $j-1$ to $j$ are opposite to each other.
Nevertheless, since we define the reflection and transmission coefficients with respect to $\lambda$, they should be the same for each interface:
\begin{equation}
    \TC{j}{j-1} = \TC{j}{j+1}, \qquad
    \RC{j}{j-1} = \RC{j}{j+1}.
\end{equation}
The same argument can be applied to the incoming wave from the region $j$ to $j-1$ (see also the right panel of Fig.~\ref{fig:cntrct}):
\begin{equation}
    \TC{j-1}{j} = \TC{j+1}{j}, \qquad
    \RC{j-1}{j} = \RC{j+1}{j}.
\end{equation}
By using these relations, we obtain
\begin{eBox}
\begin{equation}
    \label{eq:rel_jp1toj}
    \TCinv{j+1}{j} = \TC{j}{j+1} - \RC{j+1}{j} \TCinv{j+1}{j} \RC{j}{j+1}, \qquad
    \RC{j+1}{j} \TCinv{j+1}{j} = - \TCinv{j+1}{j} \RC{j}{j+1},
\end{equation}
or equivalently,
\begin{equation}
    \label{eq:rel_jm1toj}
    \TCinv{j-1}{j} = \TC{j}{j-1} - \RC{j-1}{j}
    \TCinv{j-1}{j} \RC{j}{j-1}, \qquad
    \RC{j-1}{j} \TCinv{j-1}{j} = - \TCinv{j-1}{j} \RC{j}{j-1}.
\end{equation}
\end{eBox}
Now one may rewrite the transfer matrix as
\begin{align}
	\mathbb{M}_{j-1|j} &=
        \begin{pmatrix}
            e^{-i \mathbb{k}_{z}^{(j-1)} z_{j-1|j}} & \\
            & e^{+i \mathbb{k}_{z}^{(j-1)} z_{j-1|j}}
        \end{pmatrix}
        \begin{pmatrix}
            \mathbb{1} & 
            \RC{j}{j-1} \\
            \RC{j}{j-1}  &
            \mathbb{1}
        \end{pmatrix}
        \begin{pmatrix}
            \mathbb{T}^{(j|j-1)^{-1}}  & \\
            & \mathbb{T}^{(j|j-1)^{-1}}
        \end{pmatrix}
        \begin{pmatrix}
            e^{+i \mathbb{k}_{z}^{(j)} z_{j-1|j}} & \\
            & e^{-i \mathbb{k}_{z}^{(j)} z_{j-1|j}}
        \end{pmatrix}
        ,
    \label{Mij_2} \\
    \mathbb{M}_{j|j-1} &=
    \begin{pmatrix}
        e^{-i \mathbb{k}_{z}^{(j)} z_{j-1|j}} & \\
        & e^{+i \mathbb{k}_{z}^{(j)} z_{j-1|j}}
    \end{pmatrix}
    \begin{pmatrix}
        \mathbb{1} &
        \RC{j-1}{j} \\
        \RC{j-1}{j}  &
        \mathbb{1}
    \end{pmatrix}
    \begin{pmatrix}
        \mathbb{T}^{(j-1|j)^{-1}}  & \\
        & \mathbb{T}^{(j-1|j)^{-1}} 
    \end{pmatrix}
    \begin{pmatrix}
        e^{+i \mathbb{k}_{z}^{(j-1)} z_{j-1|j}} & \\
        & e^{-i \mathbb{k}_{z}^{(j-1)} z_{j-1|j}}
    \end{pmatrix}
    .
\end{align}

\subsection{Some examples}
\label{sec:examples_pr}

Our formulation is completely independent of the concrete form of the reflection matrix $\RC{j+1}{j}$ and transmission matrix $\mathbb{T}^{(j+1|j)}$. 
Once the concrete form of them is given for an adjacent layer, we can recursively calculate any $\RC{i}{j}$ and $\mathbb{T}^{(i|j)}$.
Here we show several examples for the reflection matrix of electromagnetic waves.
See App.~\ref{app:reflection} for derivation.

\paragraph{Dielectric medium}

For the case of dielectric medium with dielectric function $\epsilon^{(i)}(\omega)$, the convenient polarization basis is TM and TE, since the reflection and transmission matrix is diagonal in this basis:
\begin{align} 
    \RC{j+1}{j} = \begin{pmatrix} R^{(j+1|j)}_{\rm TM} & 0 \\ 0 & R^{(j+1|j)}_{\rm TE}
    \end{pmatrix},~~~~~~~~~~~~
    \mathbb{T}^{(j+1|j)} = \begin{pmatrix} T^{(j+1|j)}_{\rm TM} & 0 \\ 0 & T^{(j+1|j)}_{\rm TE} 
    \end{pmatrix}.
\end{align} 
where
\begin{align} 
    R^{(j+1|j)}_{\rm TM} = \frac{\epsilon^{(j+1)}k_{z}^{(j)} - \epsilon^{(j)}k_{z}^{(j+1)}}{\epsilon^{(j+1)}k_{z}^{(j)} + \epsilon^{(j)}k_{z}^{(j+1)}},~~~~~~~~~~~~
    R^{(j+1|j)}_{\rm TE} = \frac{k_{z}^{(j)} - k_{z}^{(j+1)}}{k_{z}^{(j)} + k_{z}^{(j+1)}},
\end{align} 
and
\begin{align} 
    T^{(j+1|j)}_{\rm TM} = \frac{ 2\sqrt{\epsilon^{(j)}\epsilon^{(j+1)}}k_{z}^{(j)} }{\epsilon^{(j+1)}k_{z}^{(j)} + \epsilon^{(j)}k_{z}^{(j+1)}},~~~~~~~~~~~
    T^{(j+1|j)}_{\rm TE} = \frac{2k_{z}^{(j)}}{k_{z}^{(j)} + k_{z}^{(j+1)}},~~~~~~~~~~~~
\end{align} 
with $k_{z}^{(j)}=\sqrt{\epsilon^{(j)}(\omega)\omega^2-k_{\parallel}^2}$.

Sometimes one may want to use the helicity basis. 
In such a case the reflection and transmission matrices are given by
\begin{align} 
    &\RC{j+1}{j} =\frac{1}{2} \begin{pmatrix} 
    R^{(j+1|j)}_{\rm TM}-R^{(j+1|j)}_{\rm TE} &
    R^{(j+1|j)}_{\rm TM}+R^{(j+1|j)}_{\rm TE} 
    \\ 
    R^{(j+1|j)}_{\rm TM}+R^{(j+1|j)}_{\rm TE} &
    R^{(j+1|j)}_{\rm TM}-R^{(j+1|j)}_{\rm TE} 
    \end{pmatrix},\\
   & \mathbb{T}^{(j+1|j)} =\frac{1}{2} \begin{pmatrix} 
   T^{(j+1|j)}_{\rm TM} + T^{(j+1|j)}_{\rm TE} &
   T^{(j+1|j)}_{\rm TM} - T^{(j+1|j)}_{\rm TE} 
   \\ 
   T^{(j+1|j)}_{\rm TM} - T^{(j+1|j)}_{\rm TE} & 
   T^{(j+1|j)}_{\rm TM} + T^{(j+1|j)}_{\rm TE}
    \end{pmatrix}.
\end{align} 
It is explicitly checked that Eq.~(\ref{eq:rel_jp1toj}) is satisfied for both choices of the the basis.

\paragraph{Weyl semimetal}

Let us suppose that the medium $j$ is the vacuum and $j+1$ is the Weyl semimetal with a vector $\vec b$, which is a vector that connects two Weyl nodes in the Brillouin zone, pointing to the $z$ direction.
In this case, the convenient polarization choice is ``$+$'' and ``$-$'' mode decomposition, which corresponds to the right- and left-circular polarization in the vacuum, while it is slightly different in the Weyl semimetal~\cite{Wilson:2015wsa,Farias:2020qqp,Ema:2023kvw}. 
In this basis the reflection and transmission matrix is given by\footnote{
	Note that, in our definition of the basis, the $k_{z,+}$ ($k_{z,-}$) mode is always placed at the upper (lower) column for both the right- and left-moving mode, which means that the helicity of the upper column for the left- and right-moving ones are different. On the other hand, in the definition of Refs.~\cite{Farias:2020qqp,Ema:2023kvw}, the helicity of the upper column for the left- and right-moving modes are the same.}
\begin{align} 
    \RC{j+1}{j} = - \RC{j}{j+1} = 
    \begin{pmatrix} R_{+} & 0 \\ 0 & R_{\rm -}
    \end{pmatrix},~~~~~~~~~~~~
    \mathbb{T}^{(j+1|j)} = \begin{pmatrix} T_{+} & 0 \\ 0 & T_{\rm -} 
    \end{pmatrix},~~~~~~~~~~~~
     \mathbb{T}^{(j|j+1)} = \begin{pmatrix} T'_{+} & 0 \\ 0 & T'_{\rm -} 
    \end{pmatrix},
\end{align} 
where
\begin{align} 
    R_{\lambda} = \frac{k_{z,\lambda} - \kappa_z}{k_{z,\lambda} + \kappa_z},~~~~~~~~~~
    T_{\lambda} =\frac{2N_\lambda}{\omega}\frac{1}{k_{z,\lambda}+\kappa_z},~~~~~~~~
    T'_{\lambda} =\frac{2\omega}{N_\lambda}\frac{k_{z,\lambda} \kappa_z}{k_{z,\lambda}+\kappa_z},
    ~~~~~~~{\rm for}~~~~~~
    \lambda = \pm,
    \label{eq:RT_Weyl}
\end{align} 
with $\kappa_z = \sqrt{\omega^2-k_\parallel^2}$, $k^{2}_{z,\pm} = \kappa_z(\kappa_z \pm b)$ and 
$N_\lambda = \sqrt{\left(\kappa_z^4+\omega^2\kappa_z^2 + k_{\parallel}^2 k_{z,\lambda}^2\right)/2}$ being the normalization constant for the polarization vector.
It is also explicitly checked that they satisfy Eq.~(\ref{eq:rel_jp1toj}).

\section{Casimir energy in multilayer}
\label{sec:cas_energy_mul}
Here we derive the formula of the Casimir energy in a language developed so far, and reproduce the formula of the Casimir force known in the literature.

\subsection{Characteristic equation and argument principle}
\label{sec:ch_eq}

The appropriate boundary conditions to obtain the Casimir energy of a multilayer system shown in Fig.~\ref{fig:multilayer} are as follows: only left-moving wave in the region $0$ and only right-moving wave in the region $N+1$.
Pick up a layer $j$ with $0 < j < N+1$.
The wave functions in the region $j$ is obtained by multiplying $\mathbb{M}_{j|0}$ to the left-moving wave in the region $0$, which leads to
\begin{equation}
    \mathbb{M}_{j|0}
    \begin{pmatrix}
        0\\
        \vec{\phi}_L^{(0)}
    \end{pmatrix}
    =
    \begin{pmatrix}
        e^{-i \mathbb{k}_{z}^{(j)} z_{j-1|j}}
        \, \mathbb{R}^{(0|j)}
        \mathbb{T}^{(0|j)^{-1}} 
        e^{-i \mathbb{k}_{z}^{(0)} z_{0|1}}\,
        \vec{\phi}_{L}^{(0)} \\ 
        e^{+i \mathbb{k}_{z}^{(j)} z_{j-1|j}}
        \,\mathbb{T}^{(0|j)^{-1}}
        e^{-i \mathbb{k}_{z}^{(0)} z_{0|1}} \,
        \vec{\phi}_{L}^{(0)} 
    \end{pmatrix}.
\end{equation}
If we further multiply $\mathbb{M}_{N+1|j}$, the wave functions should become the right-moving wave so that they fulfill the boundary condition.
This requirement forces a non-trivial relation of
\begin{equation}
    e^{+i \mathbb{k}_{z}^{(j)} z_{j-1|j}}
    \,\mathbb{T}^{(0|j)^{-1}}
    e^{-i \mathbb{k}_{z}^{(0)} z_{0|1}}
    \vec{\phi}_L^{(0)}
    =
    e^{+i \mathbb{k}_z^{(j)} z_{j|j+1}}
    \,\mathbb{R}^{(N+1|j)} 
    e^{+i \mathbb{k}_z^{(j)} \Delta z_{j}}
    \mathbb{R}^{(0|j)} \mathbb{T}^{(0|j)^{-1}}
    e^{-i \mathbb{k}_{z}^{(0)} z_{0|1}}
    \vec{\phi}_L^{(0)}.
\end{equation}
To have non-trivial solution to this equation, the following characteristic equation should be satisfied
\begin{equation} \label{eq:char_Cas}
    0 = \det \qty[\mathbb{1} - e^{+i \mathbb{k}_z^{(j)} \Delta z_{j}}
    \,\mathbb{R}^{(N+1|j)} 
    e^{+i \mathbb{k}_z^{(j)} \Delta z_{j}}
    \mathbb{R}^{(0|j)} ] 
    \equiv
    f^{(0|j|N+1)} (\omega),
\end{equation}
whose solution gives the energy of mode allowed in this multilayer system. 
We label each mode by $n$ such that $\omega = \omega_n$ solves Eq.~\eqref{eq:char_Cas}.
Note that we have $ f^{(0|j|N+1)} (\omega) = f^{(N+1|j|0)} (\omega)$ by definition.

To compute the Casimir energy or force, we have to sum over all the allowed modes.
The argument principle is frequently used for this purpose as it swaps a summation over zeros of some function to an integral.
However, we cannot directly use $f^{(0|j|N+1)} (\omega)$ as an indicator of zeros.
This is because $\mathbb{R}^{(0|j)}$ or $\mathbb{R}^{(N+1|j)}$ in $f^{(0|j|N+1)} (\omega)$ involve poles as can be seen from the recursion relations given in Eqs.~\eqref{eq:recur_T} and \eqref{eq:recur_R}, which yields other contributions than the summation over zeros of $f^{(0|j|N+1)} (\omega)$ when we apply the argument principle to $f^{(0|j|N+1)} (\omega)$.

Let us take a closer look at $\mathbb{R}^{(0|j)}$ as an example to illustrate how to cancel out all the poles involved.
Pick up a layer $k$ between $0$ and $j$.\footnote{
    For $j = 1$, $\mathbb{R}^{(0|j)}$ does not involve a pole.
}
By applying the recursion relations of Eqs.~\eqref{eq:recur_T} and \eqref{eq:recur_R} to $\mathbb{R}^{(0|j)}$ for the layer $k$, one finds that the poles of $\mathbb{R}^{(0|j)}$ are indicated by the zeros of
\begin{equation} \label{eq:char_general}
    \det \qty[ 
        \mathbb{1} - e^{ + i \mathbb{k}_z^{(k)} \Delta z_k } \mathbb{R}^{(j|k)} 
        e^{ + i \mathbb{k}_z^{(k)} \Delta z_k }
        \mathbb{R}^{(l|k)}
    ]
    = f^{(l|k|j)} (\omega),
\end{equation}
for $l = 0$.
This characteristic equation involves poles in $\mathbb{R}^{(0|k)}$ or $\mathbb{R}^{(j|k)} $, indicating a recursive structure.
One has to repeat this until all the stacks of layers break up into the fundamental building block of neighboring layers.
After applying the same procedure to $\mathbb{R}^{(N+1|j)}$, we can cancel out the poles in $f^{(N+1|j|0)} (\omega)$ by multiplying all the characteristic functions involved in this procedure.
For concreteness, to cancel out the poles of $\mathbb{R}^{(0|j)}$, let us take the particular decomposition of $(0|j) \to (0|j-1) + (j-1|j)$, $(0|j-1) \to (0|j-2) + (j-2|j-1)$, $\cdots$, $(0|2) \to (0|1) + (1|2)$.
The characteristic functions involved in this process are given by
$
    \prod_{k = 1}^{j-1} f^{(0|j-k|j-k+1)} (\omega)
$.
For $\mathbb{R}^{(N+1|j)}$, we take
$(j|N+1) \to (j|j+1) + (j+1|N+1)$, $\cdots$, $(N-1|N+1) \to (N-1|N) + (N|N+1)$, which reads
$
    \prod_{k = 1}^{N-j} f^{(j+k-1|j+k|N+1)} (\omega)
$.
By multiplying these factors, we obtain the following expression useful for applying the argument principle
\begin{eBox}
\begin{equation}\label{eq:argument_simple}
    \tilde f^{(0|j|N+1)} (\omega) 
    \equiv f^{(0|j|N+1)} (\omega) 
    \qty[ \prod_{k = 1}^{j-1} f^{(0|j-k|j-k+1)} (\omega) ]
    \qty[ \prod_{k = 1}^{N-j} f^{(j+k-1|j+k|N+1)} (\omega) ],
\end{equation}
\end{eBox}
as it does not involve poles but its zeros account for all the allowed modes in this multilayer system.

Although the result seemingly depends on how we break up the stack of layers into the neighboring layers, one can guarantee its independence as follows.
By taking the determinant for the both-hand sides of Eq.~\eqref{eq:consistency_rel}, we obtain
\begin{equation}\label{eq:swap_k_l}
    f^{(i|k|j)} (\omega) f^{(i|l|k)} (\omega) = f^{(i|l|j)} (\omega) f^{(l|k|j)} (\omega),
\end{equation}
for $j > k > l > i$ or $j < k < l < i$.
The left-hand side indicates the following decomposition:
insert a layer $k$ between $j$ and $i$, and then insert a layer $l$ between $k$ and $i$, \textit{i.e.}, $(i|j) \to (i|k) + (k|j)$, $(i|k) \to (i|l) + (l|k)$.
On the other hand, the right-hand side indicates that the insertion of a layer $l$ first and then $k$, \textit{i.e.}, $(i|j) \to (i|l) + (l|j)$, $(l|j) \to (l|k) + (k|j)$.
In this way, Eq.~\eqref{eq:swap_k_l} shows that we can swap the ordering of decomposition for $k$ and $l$ while the boundaries $i$ and $j$ are fixed,  and hence the improved characteristic function \eqref{eq:argument_simple} is independent of how we break up the stacks of layers.
Furthermore, the apparent $j$-dependence of Eq.~\eqref{eq:argument_simple} is also false.
By using Eq.~\eqref{eq:swap_k_l}, we immediately find
$
    f^{(0|j|N+1)} (\omega) f^{(j|j+1|N+1)} (\omega) 
    = 
    f^{(0|j+1|N+1)} (\omega) f^{(0|j|j+1)} (\omega)
$
which implies the $j$-independence:
\begin{eBox}
\begin{equation}\label{eq:j-indep}
    \tilde f^{(0|j|N+1)} (\omega) = \tilde f^{(0|j+1|N+1)} (\omega)
    \equiv 
    \tilde f^{(0|N+1)} (\omega).
\end{equation}
\end{eBox}

\subsection{Renormalization of Casimir energy}
\label{sec:cas_E_renorm}

\begin{figure}[t]
    \centering
    \includegraphics[width=0.5\linewidth]{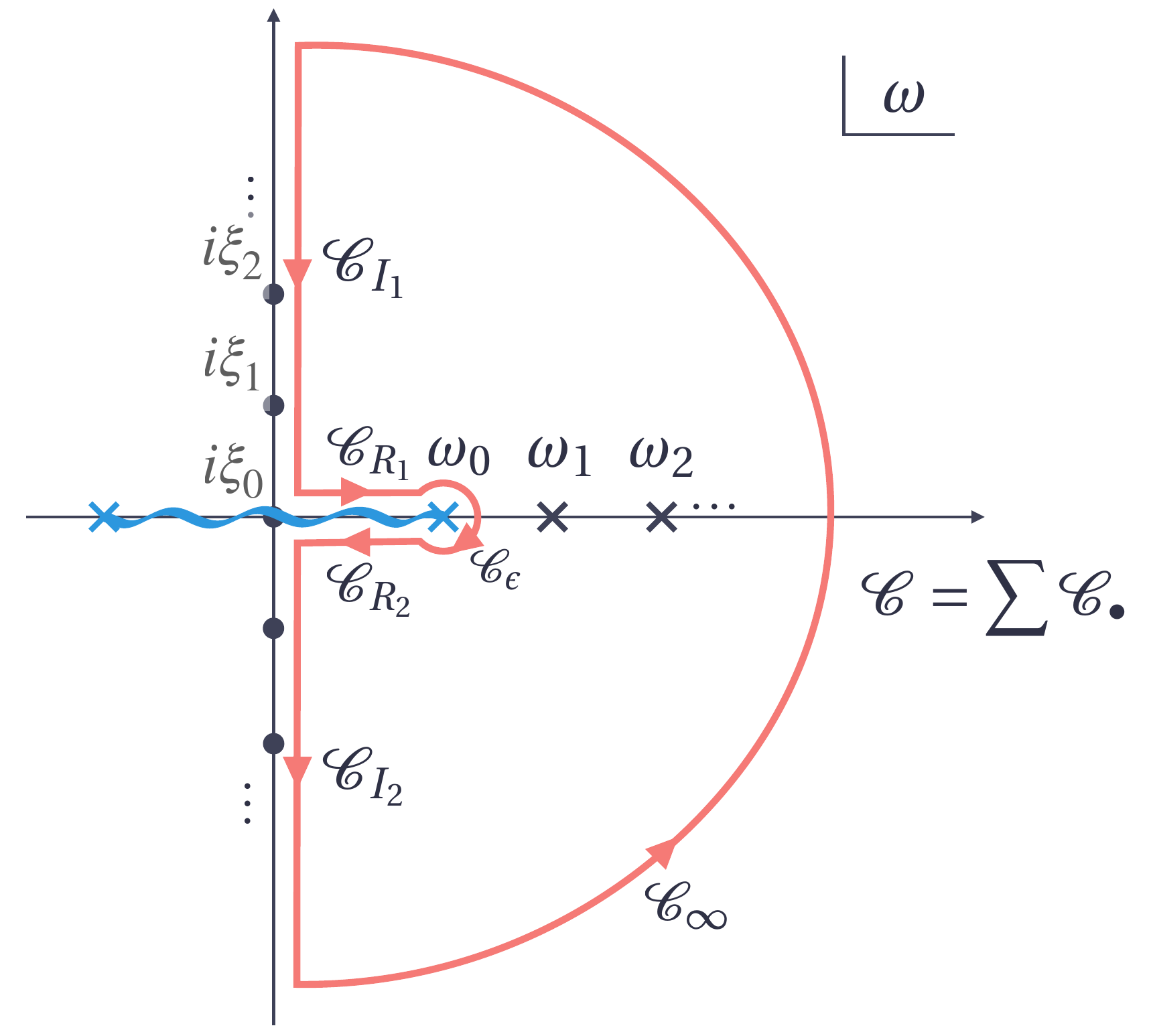}
    \caption{Contour of integral \eqref{eq:argument_principle}. Zeros of $\tilde f^{(0|j|N+1)} (\omega)$ are indicated by the crosses, $\omega_n$. Poles of $\ln (2\sinh \sfrac{\omega}{2 T})$ are given by the blobs, $i\xi_\ell$ with $\xi_\ell = 2 \pi T \ell$.
    Below the lowest allowed energy $\omega_0$, we do not have propagating modes, which leads to the branch cut indicated by a blue wavy line.
    }
    \label{fig:contour}
\end{figure}

We consider a system in thermal equilibrium with temperature $T$.
The Casimir free energy is obtained from the summation over all the allowed modes, namely
\begin{equation}
    \mathcal{F}_{\text{Cas},\Lambda}^{(0|N+1)} \equiv \sum_\lambda {\sum_{n \geqslant 0}}' \int_{{\bm k}_\parallel} 
    \qty[
        \frac{\omega_n}{2} + T \ln \qty( 1 - e^{- \omega_n / T} )
    ] W ( \omega_n/\Lambda )
    = 
    \sum_\lambda
    {\sum_{n \geqslant 0}}' \int_{{\bm k}_\parallel} 
    T \ln \qty( 2 \sinh \frac{\omega_n}{2 T} )
    W (\omega_n/\Lambda), 
\label{Fcas}
\end{equation}
with the allowed frequencies $\omega_n$ being determined by the zeros of $\tilde f^{(0 | j | N+1)} ( \omega )$.
The prime on the summation implies that the summation of $n = 0$ should be multiplied by a factor of $\sfrac{1}{2}$.
We used shorthand notation $ \int_{{\bm k}_\parallel} = \int\frac{\dd^2{\bm k}_\parallel}{(2\pi)^2}$.
This expression is however divergent and hence we introduce
a regulator denoted by $W(z)$ that goes to zero sufficiently smooth and fast for $|z| \to \infty$.
Physically, there should be a counter term to obtain finite vacuum energy in the original Lagrangian.
The renormalized Casimir energy is defined by
\begin{align}
    \mathcal{F}_\text{Cas,r}^{(0|N+1)} = \lim_{\Lambda \to \infty} \qty( \mathcal{F}_{\text{Cas},\Lambda}^{(0|N+1)} - \mathcal{F}_{\text{vac},\Lambda} ),
    \label{Fcas_r}
\end{align}
where $\mathcal F_{\text{vac},\Lambda}$ is the vacuum energy, which is given by
\begin{equation}
    \mathcal{F}_{\text{vac},\Lambda} = L_z \sum_\lambda \int\frac{\dd k_z}{2\pi} \int_{{\bm k}_\parallel} 
    \frac{\omega}{2} W ( \omega/\Lambda ).
\label{Fvac}
\end{equation}
where $L_z = \Delta z_{(N|1)}$ is the total width of the system.
With this counter term, the Casimir energy is always finite.

However, when it comes to actual computation, an appropriate subtraction of vacuum energy could be involved.
In most cases, we are dealing with a low-energy effective field theory rather than the original microscopic Lagrangian, for instance examples given in Sec.~\ref{sec:examples_pr}. Such a low-energy effective field theory should be associated with a physical cutoff scale, which is specified once its construction from the original Lagrangian is given.
However, if we naively compute the Casimir energy within the low-energy description without the information of physical cutoff scale, we end up with fictitious UV divergences.
They should be renormalized to give finite results based on the physical cutoff scale, or in other words, we should have appropriate counter terms once its construction from the original microscopic theory is specified.
In this sense, the computation of renormalized Casimir energy sometimes requires the microscopic details more than the wave equation in low-energy effective field theory \eqref{eq:phi_eom}.

Nevertheless, if we restrict ourselves to the Casimir force in a realistic system, such UV dependent terms do not contribute.
This is because such terms are associated with the Casimir energy of the medium itself.
Hence they are irrelevant unless we can really control the size of the medium, which is practically not possible.\footnote{
    Such contributions might be relevant in a more particle-physics-oriented context, where the system is separated by a domain wall or a brane, and the size of the individual domain can change by the dynamics of the domain wall or the brane.
    }
The layers with variable $\Delta z_j$ are usually vacuum, and hence the UV dependent terms are absent in the Casimir force.
We can also consider the case where the layers with $\Delta z_j$ are non-trivial medium, but in this case, the whole system is emersed with the same medium, and the entire size of the medium is fixed.
This results in the cancellation of the UV dependent terms for the Casimir force because the Casimir energy does not change by the variation of $\Delta z_j$ in the end.
For this reason, we only discuss the UV independent contribution to the vacuum energy and drop the UV dependent terms in the following.
We will come back to this point in the end of Sec.~\ref{sec:cas_F}.

We utilize the argument principle to convert the summation over the allowed modes $\omega_n$ to an integral over $\omega$.
By taking the contour integral shown in Fig.~\ref{fig:contour}, one may rewrite the summation over $n \geqslant 1$ as follows 
\begin{align}
    \sum_\lambda \sum_{n \geqslant 1}
    T \ln \qty( 2 \sinh \frac{\omega_n}{2 T} ) W (\omega_n/\Lambda)
    &= 
    \frac{1}{2 \pi i} \oint_{\cal C}
    T \ln \qty( 2 \sinh \frac{\omega}{2 T} ) W(\omega/\Lambda) ~
    \dd \ln \tilde f^{(0|N+1)} (\omega),
    \label{eq:argument_principle}
\end{align}
where the $\sinh$ function involves poles in the imaginary axis as $\omega = i \xi_\ell$ with $\xi_\ell \equiv 2 \pi T \ell$.
The integral on the arc $\mathcal{C}_\infty$ vanishes in the limit of infinite diameter due to the regulator $W(|\omega|/\Lambda)$.
The integral on the infinitesimal arc $\mathcal{C}_\epsilon$, on the other hand, gives
\begin{equation}
    \frac{1}{2 \pi i} \int_{{\cal C}_\epsilon}
    T \ln \qty( 2 \sinh \frac{\omega_0}{2 T} ) ~
    \sum_\lambda\frac{\dd \omega}{2 \qty(\omega - \omega_0)}
    = - \sum_\lambda \frac{1}{2}  T \ln \qty( 2 \sinh \frac{\omega_0}{2 T} ),
\end{equation}
which cancels out the remaining contribution of $n = 0$.
Here we assume that the lowest allowed energy $\omega_0$ is common for all the polarizations.
Thus we have
\begin{align}
	\mathcal{F}_{\text{Cas},\Lambda}^{(0|N+1)} = \frac{1}{2 \pi i} \int_{{\cal C}_{I_1+I_2}+{\cal C}_{R_1+R_2}}
    T \ln \qty( 2 \sinh \frac{\omega}{2 T} ) W(\omega/\Lambda) ~
    \dd \ln \tilde f^{(0|N+1)} (\omega).
\end{align}
In the following we evaluate this integral.
First we carefully quantify the behavior of complex function in each contour ${\cal C}_{I_1+I_2}$ and ${\cal C}_{R_1+R_2}$. Then we extract UV divergent contributions and see how it is renormalized. 
Finally we will determine the finite expression for the renormalized Casimir energy.

\subsubsection*{Some preliminaries}

Below $\omega_0$, there are no propagating modes, which leads to the branch cut in general.
Going around the branch point, we expect that the transverse momentum is flipped.
For instance, in the case of dielectric material, the dispersion relation, $k_z = \sqrt{\epsilon \omega^2 - k_\parallel^2}$, implies $k_z = i\hat k_z$ on $\mathcal{C}_{R_1}$ while $k_z = - i \hat k_z$ on $\mathcal{C}_{R_2}$ with $\hat k_z \equiv \sqrt{ k_\parallel^2 - \epsilon \omega^2}$.
Also, in some cases, the polarizations can be exchanged.
For instance, in the case of Weyl semimetals, where the dispersion relation is $k_{z\pm} = \sqrt{\kappa_z (\kappa_z \pm b )}$ with $\kappa_z = \sqrt{\omega^2 - k_\parallel^2}$, we expect $k_{z\pm} = i \hat k_{z\pm}$ on $\mathcal{C}_{R_1}$ while $k_{z\pm} = - i \hat k_{z \mp}$ on $\mathcal{C}_{R_2}$ with $\hat\kappa_z = \sqrt{k_\parallel^2 - \omega^2}$ and $\hat k_{z \pm} \equiv \sqrt{\hat\kappa_z (\hat\kappa_z \mp i b )}$~\cite{Wilson:2015wsa,Ema:2023kvw}.
Motivated by these observations, we assume the following property for the transverse momentum throughout this paper:
\begin{eBox}
\begin{equation}
    \mathbb{k}_z ~ \text{on the real axis with $\omega > \omega_0$} 
    ~\longmapsto~
    + i \hat{\mathbb{k}}_z ~\text{on $\mathcal{C}_{R_1+I_1}$} 
    ~\longmapsto~
    + i \check{\mathbb{k}}_z =
    -i \mathbb{P} \hat{\mathbb{k}}_z \mathbb{P}^{-1}
    \equiv -i \hat{\mathbb{k}}_{z P} ~\text{on $\mathcal{C}_{R_2 + I_2}$},
    \label{eq:chr_highE}
\end{equation}
\end{eBox}
where the possible exchange of polarization basis is denoted by the parity transformation $\mathbb{P}$ with respect to the $z$-axis.
This also induces a non-trivial transformation of the reflected/transmitted waves:
\begin{eBox}
\begin{align}
    &\RC{j}{j+1} ~ \text{on the real axis with $\omega > \omega_0$} 
    ~\longmapsto~
    \tilRC{j}{j+1}
    ~\text{on $\mathcal{C}_{R_1+I_1}$} 
    ~\longmapsto~
    \hatRC{j}{j+1} =
    \mathbb{P}\tilRC{j}{j+1} \mathbb{P}^{-1} \equiv
    \tilRC{j}{j+1}_P
    ~\text{on $\mathcal{C}_{R_2+I_2}$} \label{eq:analytic_c_R} \\
    &\TC{j}{j+1} ~ \text{on the real axis with $\omega > \omega_0$} 
    ~\longmapsto~
    \tilTC{j}{j+1}
    ~\text{on $\mathcal{C}_{R_1+I_1}$} 
    ~\longmapsto~
    \hatTC{j}{j+1} =
    \mathbb{P} \tilTC{j}{j+1} \mathbb{P}^{-1} \equiv
    \tilTC{j}{j+1}_P
    ~\text{on $\mathcal{C}_{R_2+I_2}$}. \label{eq:analytic_c_T}
\end{align}
\end{eBox}

In the following, to make our discussion as simple as possible, we consider the case where the both ends, \textit{i.e.}, the region $0$ and $N+1$, are perfect conductors.\footnote{
	One might think that this is a special setup. However, practically it does not reduce generality since one can effectively reproduce any setup by taking these perfect conductors infinitely far from the remaining layers. See Sec.~\ref{sec:cas_F}.
}
By using relations \eqref{eq:rel_jm1toj}, one may rewrite the recursion relation Eq.~\eqref{eq:recur_R} as follows:
\begin{equation}
    \label{eq:R0j_new}
    \begin{split}
    \RC{0}{j} =&
    \Big[
        - \TCinv{j-1}{j} \RC{j}{j-1} e^{-i \mathbb{k}_{z}^{(j-1)} \Delta z_{j-1}}
        +
        \TCinv{j-1}{j} e^{i \mathbb{k}_{z}^{(j-1)} \Delta z_{j-1}} \RC{0}{j-1}
    \Big] \\[-.2em]
    & \times
    \Big[
        \TCinv{j-1}{j} e^{- i \mathbb{k}_{z}^{(j-1)} \Delta z_{j-1}} - \TCinv{j-1}{j} \RC{j}{j-1} e^{i \mathbb{k}_{z}^{(j-1)} \Delta z_{j-1}} \RC{0}{j-1} 
    \Big]^{-1} \,.
    \end{split}
\end{equation}
Utilizing Eqs.~\eqref{eq:analytic_c_R} and \eqref{eq:analytic_c_T}, one can show
\begin{eBox}
\begin{equation}
    \RC{0}{j} ~ \text{on the real axis with $\omega > \omega_0$}
    ~\longmapsto~
    \tilRC{0}{j} ~\text{on $\mathcal{C}_{R_1+I_1}$}
    ~\longmapsto~
    \hatRC{0}{j} =
    \tilRCinv{0}{j}_P ~\text{on $\mathcal{C}_{R_2+I_2}$}.
    \label{eq:analytic_c_R0j}
\end{equation}
\end{eBox}
The last relation can be shown by induction as follows.
For $j=1$, the perfect conductor at $z \leqslant z_{0|1}$ implies $\tilRC{0}{1} = \mathbb{1}$ and $\hatRC{0}{1} = \mathbb{1}$, which is consistent with $\hatRC{0}{1} = \tilRCinv{0}{1}_P$.
Then, assuming that the relation $\hatRC{0}{k} = \tilRCinv{0}{k}_P$ holds for $k \leqslant j-1$, one can show
\begin{align}
    \hatRC{0}{j}
    =&
    \Big[
        - \tilTCinv{j-1}{j}_P \tilRC{j}{j-1}_P e^{- \hat{\mathbb{k}}_{z P}^{(j-1)} \Delta z_{j-1}}
        + \tilTCinv{j-1}{j}_P e^{- \hat{\mathbb{k}}_{z P}^{(j-1)} \Delta z_{j-1}} \tilRCinv{0}{j-1}_P
    \Big] \nonumber \\[-.2em]
    & \times
    \Big[
        \tilTCinv{j-1}{j}_P e^{- \hat{\mathbb{k}}_{z P}^{(j-1)} \Delta z_{j-1}} - \tilTCinv{j-1}{j}_P \tilRC{j}{j-1} e^{\hat{\mathbb{k}}_{z P}^{(j-1)} \Delta z_{j-1}} \tilRCinv{0}{j-1}_P 
    \Big]^{-1}\nonumber \\
    =&\tilRCinv{0}{j}_P,
    \label{eq:hatRC0j}
\end{align}
which completes the proof.
The same relation holds for the other side of the boundary, \textit{i.e.}, $\hatRC{N+1}{j} = \tilRCinv{N+1}{j}_P$.

Now we are ready to extract the UV divergent contribution.
As shown in Sec.~\ref{sec:ch_eq}, the characteristic function is independent of how we break up the stacks of layers, which allows us to take the following form:
\begin{equation}
    \label{eq:chr_simple}
    \tilde f^{(0|N+1)} (\omega) 
    = f^{(0|1|2)} (\omega) f^{(0|2|3)} (\omega) \cdots
    f^{(0|N|N+1)} (\omega).
\end{equation}
By using the relations given in Eqs.~\eqref{eq:analytic_c_R}, \eqref{eq:analytic_c_T}, and \eqref{eq:analytic_c_R0j}, we obtain the following relation:
\begin{equation}
    \label{eq:chr_relation}
    \check{f}^{(0|N|N+1)} \check{f}^{(0|N - 1|N)} \cdots \check{f}^{(0|j-1|j)} = \hat{f}^{(0|N|N+1)} \hat{f}^{(0|N - 1|N)} \cdots \hat{f}^{(0|j-1|j)} \det \Bigg[ \prod_{k=j-1}^{N}e^{2 \hat{\mathbb{k}}_z^{(k)} \Delta z_k } \Bigg] 
    \det \Bigg[ \tilRCinv{0}{j-1} \Bigg].
\end{equation}
This can also be shown by induction.
For $j=N+1$, one may confirm that
\begin{equation}
    \check{f}^{(0|N|N+1)} = \det \qty[ \mathbb{1} - e^{ 2 \hat{\mathbb{k}}_z^{(N)} \Delta z_{N} }
    \tilRCinv{0}{N}  ]
    = \hat{f}^{(0|N|N+1)} \det \qty[e^{ 2 \hat{\mathbb{k}}_z^{(N)} \Delta z_{N} }]
    \det \qty[ \tilRCinv{0}{N}  ].
\end{equation}
Then, assuming that the relation holds for $j = k + 1$, one can show
\begin{align}
    \check{f}^{(0|N|N+1)} \check{f}^{(0|N - 1|N)} \cdots \check{f}^{(0|k-1|k)}
    &= \hat{f}^{(0|N|N+1)} \hat{f}^{(0|N - 1|N)} \cdots \hat{f}^{(0|k|k+1)} \det \Bigg[ \prod_{\ell =k}^{N} e^{2 \hat{\mathbb{k}}_z^{(\ell)} \Delta z_\ell } \Bigg] \nonumber \\
    & \quad \times \det \Bigg[ \tilRCinv{0}{k} \Bigg]
    \det \Bigg[ \mathbb{1} - e^{\hat{\mathbb{k}}_z^{(k-1)}\Delta z_{k-1}} \tilRC{k}{k-1} e^{\hat{\mathbb{k}}_z^{(k-1)}\Delta z_{k-1}} \tilRCinv{0}{k-1} \Bigg] \label{eq:chr_recur1} \\
    & = \hat{f}^{(0|N|N+1)} \hat{f}^{(0|N - 1|N)} \cdots \hat{f}^{(0|k|k+1)} \det \Bigg[ \prod_{\ell =k}^{N} e^{2 \hat{\mathbb{k}}_z^{(\ell)} \Delta z_\ell } \Bigg] \hat{f}^{(0|k-1|k)} \nonumber \\
    &\quad \times \det \Bigg[ e^{2 \hat{\mathbb{k}}_z^{(k-1)} \Delta z_{k-1} } \Bigg] \det \Bigg[ \tilRCinv{0}{k-1} \Bigg].
    \label{eq:chr_recur2}
\end{align}
Here we have used Eq.~\eqref{eq:R0j_new} to obtain the last line.
Therefore, by combining Eqs.~\eqref{eq:chr_simple} and \eqref{eq:chr_relation}, we eventually arrive at
\begin{eBox}
\begin{equation}
    \label{eq:chr_UVdiv}
    \hat{\tilde{f}}^{(0|N+1)} (\omega) ~\text{on $\mathcal{C}_{R_1+I_1}$}
    ~\longmapsto~
    \check{\tilde{f}}^{(0|N+1)} (\omega)
    = \hat{\tilde{f}}^{(0|N+1)} (\omega)\,
    \det \qty[ \prod_{k = 1}^{N}  e^{2 \hat{\mathbb{k}}_z^{(k)} \Delta z_{k} } ] ~\text{on $\mathcal{C}_{R_2+I_2}$}.
\end{equation}
\end{eBox}

\subsubsection*{Structure of UV divergence}

Noticing that $\ln \hat{\tilde{f}}^{(0|N+1)}$ rapidly approaches to zero on the arc $\mathcal{C}_\infty$, one can see from Eq.~\eqref{eq:chr_UVdiv} that the UV divergence resides in the contributions on the contour $\mathcal{C}_{R_2} + \mathcal{C}_{I_2}$, \textit{i.e.,} th exponential factor in the determinant of Eq.~\eqref{eq:chr_UVdiv}.
The UV divergence in Eq.~\eqref{eq:argument_principle} is originated from
\begin{equation}
    \ln \det \qty[
        \prod_{k = 1}^{N}  e^{2 \hat{\mathbb{k}}_z^{(k)} \Delta z_{k} }
    ]
    = \tr \qty[ \sum_{k = 1}^{N}
        2 \hat{\mathbb{k}}_z^{(k)} \Delta z_{k} ].
\end{equation}
One may estimate the strongest UV divergence in Eq.~\eqref{eq:argument_principle} as follows
\begin{align}
    \mathcal{F}^{(0|N+1)}_{\text{Cas},\Lambda}
    &\supset
    \int_{\bm{k}_\parallel}\int_{\mathcal{C}_{R_2 + I_2}} \frac{ \dd \omega}{2 \pi i}\, \frac{\omega}{2}\, \tr \qty[ \sum_{k=1}^{N} 2 \frac{\partial \hat{\mathbb{k}}_z^{(k)}}{\partial \omega} \Delta z_k ]\, W(\omega/\Lambda) \nonumber \\[.5em]
    &\qquad \xlongrightarrow{\text{UV limit of}~ \omega,\bm{k}_\parallel} ~
    2 L_z \sum_\lambda
    \int_{\bm{k}_\parallel}\int_{\mathcal{C}_{R_2 + I_2}} \frac{ \dd \omega}{2 \pi i}\, \frac{\omega}{2}\, \frac{-\omega}{\sqrt{k_\parallel^2 - \omega^2}} \, W(\omega/\Lambda).
    \label{eq:strong_UV}
\end{align}
Here we have taken the UV limit so that $\hat{\mathbb{k}}_z^{(\bullet)} \rightarrow \sqrt{k_\parallel^2 - \omega^2}$, rewritten $\Delta z_{(N|1)} = L_z$, and used $\tr [\mathbb{1}] = \sum_\lambda$.
By changing the contour $\mathcal{C}_{R_2 + I_2}$ to the real axis, we obtain
\begin{align}
    \text{(r.h.s. of Eq.~\eqref{eq:strong_UV})}
    &=  2 L_z \sum_\lambda
    \int_{\bm{k}_\parallel}\int_{k_{\parallel}}^\infty \frac{ \dd \omega}{2 \pi}\, \frac{\omega}{2}\, \frac{\omega}{\sqrt{\omega^2 - k_\parallel^2}} \, W(\omega/\Lambda) \\
    &=
    L_z \sum_\lambda  \int_{\bm{k}_\parallel} \int_{-\infty}^\infty \frac{\dd k_z}{2 \pi} \left.\frac{\omega}{2} W (\omega/\Lambda) \right|_{\omega = \sqrt{k_z^2 + k_\parallel^2}}
    = \mathcal{F}_{\text{vac},\Lambda}.
    \label{eq:strong_UV2}
\end{align}
It is clear that the last expression given in Eq.~\eqref{eq:strong_UV2} coincides with the divergence associated with the cosmological constant at vacuum $\propto \Lambda^4$, given in Eq.~\eqref{Fvac}.

Although we have focused on the strongest UV divergence, the first line of Eq.~\eqref{eq:strong_UV} implies that the UV divergences are generally associated with lower order divergences such as $\Lambda^2 b^2$ and $b^4 \log \Lambda$ with a certain mass-dimension scale of medium being $b$ (under the assumption of invariance w.r.t.\ $b \to - b$).\footnote{
	Here $\Lambda$ may not be the cutoff scale of the elementary quantum field theory but that of the effective field theory of the material. 
}
The renormalization of these terms strongly depends on how we construct the multilayer system, and a finite term such as $\propto b^4$ would remain after the renormalization.
Similarly, for dielectric material with dielectric constant $\epsilon$, there would also appear nontrivial UV dependent term where the UV scale corresponds to the scale at which $\epsilon$ becomes sufficiently close to unity.
These divergences, however, do not lead to measurable Casimir force  in physically reasonable setups as shown in the next subsection.

\begin{figure}[t]
    \centering
    \includegraphics[width=0.5\linewidth]{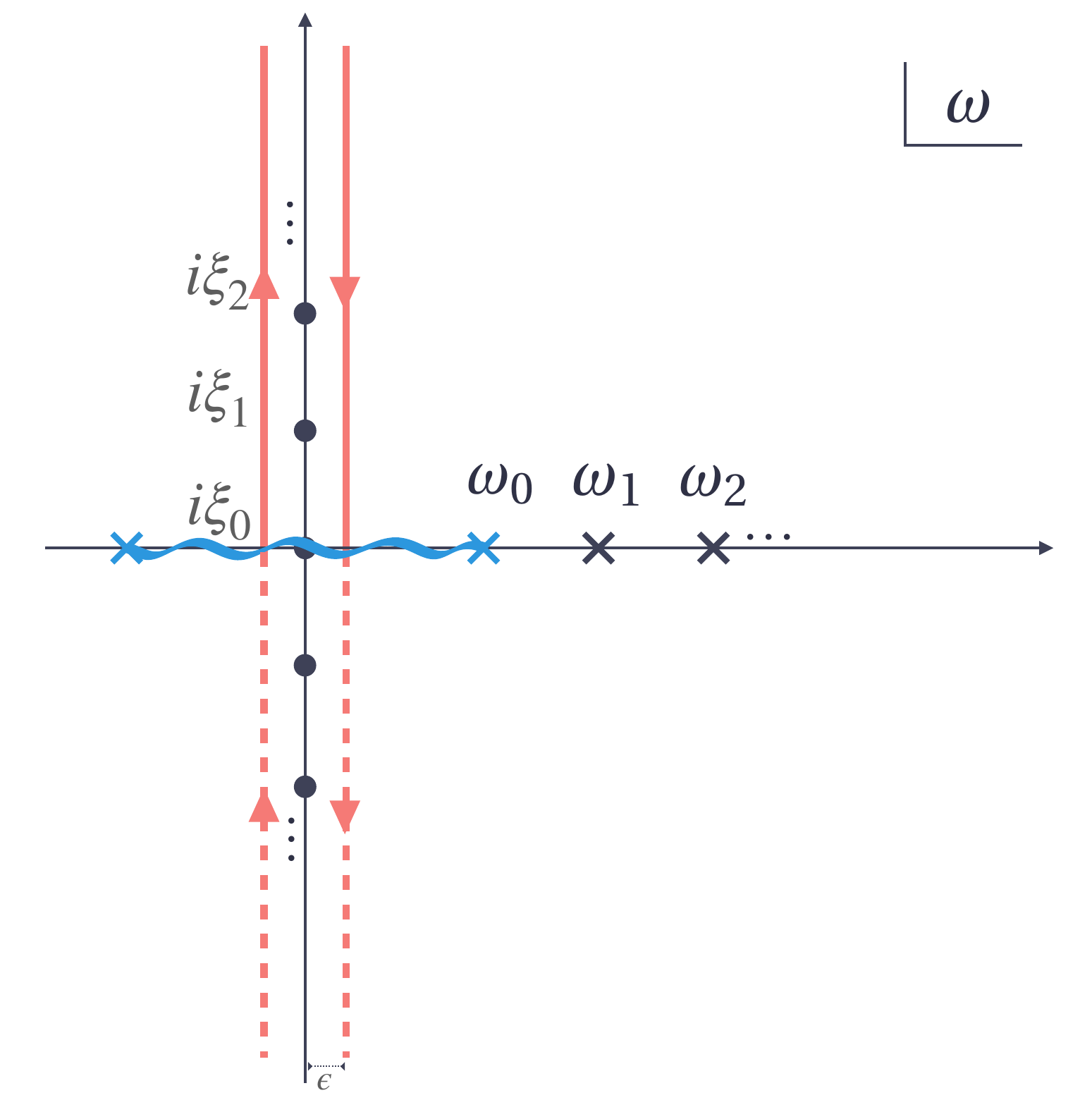}
    \caption{
        The contour for the integration in Eq.~\eqref{eq:contour_imaginary}.
        After extracting the UV divergence, the integration of $\hat{\tilde{f}}$ on the entire imaginary axis $\mathcal{C}_{I_1 + I_2}$ can be regarded as the integration from $i \infty + \epsilon$ to $-i \infty +\epsilon$ across the branch cut.
        Using $\hat{\tilde{f}} (\omega) = \hat{\tilde{f}} (-\omega)$, one can show that the integration from $i \infty + \epsilon$ to $-i \infty +\epsilon$ is equal to the integration from $-i \infty - \epsilon$ to $i \infty -\epsilon$.
        This back-and-forth integral is converted into the summation over the Matsubara frequencies $\xi_\ell$ by means of the Cauchy theorem.
    }
    \label{fig:contour2}
\end{figure}

\subsubsection*{Renormalized Casimir energy}

Finally, we move on to the finite contribution in Eq.~\eqref{eq:argument_principle}.
The characteristic function $\hat{\tilde{f}}^{(0|N+1)}$ is an even function in $\omega$.
This implies that not only the back-and-forth integral on the real axis $\mathcal{C}_{R_1 + R_2}$ cancels out, but also the integral on the imaginary axis $\mathcal{C}_{I_1 + I_2}$ can be written as
\begin{align}
    \frac{1}{2 \pi i} \int_{i \infty + \epsilon}^{- i \infty + \epsilon} T \ln \qty( \sinh \frac{\omega}{2 T} ) \, \dd \ln \hat{\tilde{f}}^{(0|N+1)} (\omega) 
    &=
    - \frac{1}{2\pi i}\int_{i \infty + \epsilon}^{-i \infty + \epsilon} \dd \omega\, \coth \qty(\frac{\omega}{2 T})\, \ln \hat{\tilde{f}}^{(0|N+1)} (\omega) \\[.5em]
    &=
    - \frac{1}{4\pi i} \qty( \int_{i \infty + \epsilon}^{-i \infty + \epsilon} + \int_{-i \infty - \epsilon}^{i \infty - \epsilon} )\, \dd \omega\, \coth \qty(\frac{\omega}{2 T})\, \ln \hat{\tilde{f}}^{(0|N+1)} (\omega) \label{eq:contour_imaginary} \\[.5em]
    &=
    T {\sum_{\ell \geqslant 0}}' \ln \hat{\tilde{f}}^{(0|N+1)} (i \xi_\ell),
\end{align}
where we have dropped $W(\omega/\Lambda)$ as this integral is finite, performed the integration by parts in the first line, and used the even property of $\hat{\tilde{f}}^{(0|N+1)}$ in the second line.
Thus, the summation over the allowed modes $\omega_n$ is converted into the Matsubara frequencies $\xi_\ell$.
See also Fig.~\ref{fig:contour2} for the contour of the integration in Eq.~\eqref{eq:contour_imaginary}.

Summing up all the contributions, we obtain the following expression for the renormalized Casimir energy
\begin{eBox}
\begin{equation}
    \mathcal{F}^{(0|N+1)}_\text{Cas,r} =   T {\sum_{\ell \geqslant 0}}' \int_{{\bm k}_\parallel} \ln \hat{\tilde{f}}^{(0|N+1)} (i \xi_\ell) + \Delta \mathcal{F}^{(0|N+1)}_{\text{Cas,r}},
\end{equation}
\end{eBox}
where a possible finite term remaining after the renormalization is denoted by $\Delta \mathcal{F}^{(0|N+1)}_{\text{Cas,r}}$, which strongly depends on the UV details of the system as mentioned earlier.
In the zero temperature limit, one should make a replacement $T\sum_{\ell\geqslant 0}'\to \int_0^\infty \frac{\dd\xi}{2\pi}$.
Its independence of the inserted layer $j$ is guaranteed by Eq.~\eqref{eq:j-indep}.
For concreteness, let us adopt a particular expression $\hat{\tilde{f}}^{(0|j|N+1)}$ given in Eq.~\eqref{eq:argument_simple} as a representative.
Then the renormalized Casimir energy can be expressed as
\begin{equation}\label{eq:Cas_energy_simple}
    \mathcal{F}_\text{Cas,r}^{(0|N+1)}  = 
    T {\sum_{\ell \geqslant 0}}' \int_{{\bm k}_\parallel}
    \qty[
    \ln \hat{f}^{(0|j|N+1)} (i \xi_\ell) +
    \sum_{k = 1}^{j-1} \ln \hat{f}^{(0|k|k+1)} (i \xi_\ell) +
    \sum_{k = 1}^{N-j} \ln \hat{f}^{(N-k|N-k+1|N+1)} (i \xi_\ell) ].
\end{equation}
Practically this expression is convenient for the calculation of the Casimir force acting on the layer $j$, as shown in the following subsection.

\subsection{Formula of Casimir force}
\label{sec:cas_F}

The Casimir force acting on the boundary of a layer $j$ is given by taking derivative of $\mathcal{F}_\text{Cas,r}$ with respect to $\Delta z_j$ while all $\Delta z_k$ with $k \neq j$ are fixed, \textit{i.e.},\footnote{
    The perfect conductors at the $0$ and $N+1$ layers should also have some fixed width and hence the outer region of the perfect conductors are assumed to be vacuum.
}
\begin{equation}\label{eq:F_Cas_j}
    F^{(0|j|N+1)}_\text{Cas} \equiv - \qty( \frac{\partial \mathcal{F}_\text{Cas,r}^{(0|N+1)} }{\partial \Delta z_j} )_{\Delta z_{k\neq j}}.
\end{equation}
For this purpose, a particular expression of $\hat{\tilde{f}}^{(0|N+1) }( \omega )$ is useful, where the layer $j$ is inserted between $N+1$ and $0$, \textit{i.e.}, $\hat{\tilde{f}}^{(0|j|N+1)} (\omega)$ (as done in Eq.~(\ref{eq:Cas_energy_simple})).
By definition given in Eq.~\eqref{eq:char_general}, the characteristic function of $f^{(j|k|l)}(\omega)$ only depends on $\Delta z_k$.
Therefore, the $\Delta z_j$ dependence is extracted by $\hat{f}^{(0|j|N+1)} (\omega)$ while the others multiplied to cancel out the poles of $\hat{f}^{(0|j|N+1)} (\omega)$ never depend on $\Delta z_j$,\textit{i.e.,}
\begin{equation}
    F^{(0|j|N+1)}_\text{Cas}
    = 
    - T {\sum_{\ell \geqslant 0}}' \int_{{\bm k}_\parallel} 
    \frac{\partial \ln \hat{\tilde{f}}^{(0|j|N+1)} (i \xi_\ell)}{\partial \Delta z_j}
    = 
    - T {\sum_{\ell \geqslant 0}}' \int_{{\bm k}_\parallel}
    \frac{\partial \ln  \hat{f}^{(0|j|N+1)} (i \xi_\ell)}{\partial \Delta z_j}.
\end{equation}
The Casimir force acting on a layer $j$ now reads
\begin{eBox}
\begin{equation}
    \label{eq:F_cas_j_formula} 
    F^{(0|j|N+1)}_\text{Cas}
    =
    - T {\sum_{\ell \geqslant 0}}' \int_{{\bm k}_\parallel}
    \tr \qty[
        e^{- \hat{\mathbb{k}}_{z, \ell}^{(j)} \Delta z_{j}} \,
        \qty{ \hat{\mathbb{k}}_{z, \ell}^{(j)} , \tilRC{N+1}{j} } e^{- \hat{\mathbb{k}}_{z,\ell}^{(j)} \Delta z_{j}} \tilRC{0}{j}\,
        \qty(\mathbb{1} - e^{- \hat{\mathbb{k}}_{z,\ell}^{(j)} \Delta z_{j}} \,\tilRC{N+1}{j} e^{- \hat{\mathbb{k}}_{z,\ell}^{(j)} \Delta z_{j}} \tilRC{0}{j} )^{-1}
        ],
\end{equation}
\end{eBox}
where we define $\hat{\mathbb{k}}_{z, \ell}^{(j)} \equiv \hat{\mathbb{k}}_z^{(j)} (i \xi_\ell)$ and the anti-commutator is denoted by $\{ \bullet, \bullet \}$.

Assuming that a particular basis of polarization diagonalizes all the matrices appearing above, one may reproduce the well-known formula
\begin{equation}
 \label{eq:F_cas_j_formula_diagonal} 
    F_\text{Cas}^{(0|j|N+1)}  = - T \sum_{\lambda} {\sum_{\ell \geqslant 0}}' \int_{{\bm k}_\parallel}
    \frac{ 2 \hat{k}_{z, \ell\lambda}^{(j)}  e^{- 2 \hat{k}_{z,\ell\lambda}^{(j)} \Delta z_{j}} \,
    \hat{R}_\lambda^{(N+1|j)} \hat{R}_\lambda^{(0|j)}}{1 - e^{- \hat{k}_{z,\ell\lambda}^{(j)} \Delta z_{j}} \,\hat{R}^{(N+1|j)}_\lambda e^{- \hat{k}_{z,\ell\lambda}^{(j)} \Delta z_{j}} \hat{R}_\lambda^{(0|j)}}.
\end{equation}

Note that, in deriving (\ref{eq:F_Cas_j}), we have dropped the $\Delta \mathcal{F}^{(0|N+1)}_{\text{Cas,r}}$ term in the Casimir energy.
It is justified as follows.
The first line of (\ref{eq:strong_UV}) explicitly indicates that $\Delta \mathcal{F}^{(0|N+1)}_{\text{Cas,r}}\sim \sum_k\left({\rm \Lambda ~dependent~term}\right)\times\Delta z_k$, where the $\Lambda$ dependent part is just determined by the local information of each layer, whatever the concrete form is.
Therefore, as far as $\Delta z_k$ is fixed quantity, rather than variable quantity, it does not lead to Casimir force.
Physically, changing the width $\Delta z_k$ of the layer $k$ with some nontrivial UV structure, while the width of all other layers are fixed, induces infinite (or at least UV cutoff dependent) change of energy.
Such a setup is unlikely.
In other words, if the width $\Delta z_k$ changes, it should be compensated by the change of other layer's width.
Thus the assumption made in (\ref{eq:F_Cas_j}) that the width of all other layers are fixed is not justified in such a setup.\footnote{	Ref.~\cite{Fukushima:2019sjn} calculated Casimir force acting on a chiral medium sandwiched by two metal plates. It leads to a divergent Casimir force with an implicit assumption that the outer regions of the metal plates are vacuum. If the outer regions are also filled by the same chiral medium, such a divergence would not appear. In our formulation, the origin and physical meaning of such a (fictious) divergence are clear.}

\subsubsection*{Example}

Now we give an example for the Casimir force with simple setup.
Notice that, in deriving the formula for the Casimir force~(\ref{eq:F_cas_j_formula}), we supposed that the leftmost and rightmost layers (layer $0$ and $N+1$) are perfect conductors.
In many applications, however, one may often want to consider more general setups in which the leftmost and rightmost are not necessarily perfect conductors.
Here we show that our formula can be applied to even such cases just by taking the width of the layer $1$ and $N$ to be very large.
Assuming $\Delta z_1, \Delta z_N \to \infty$ in Eqs.~(\ref{eq:recur_T}) and (\ref{eq:recur_R}),
we easily find $\tilRC{0}{j} \simeq \tilRC{1}{j}$ and $\tilRC{N+1}{j} \simeq \tilRC{N}{j}$.
In this way, we can effectively ``forget'' about the layer $0$ and $N+1$ and calculate the Casimir force acting on any body in any medium.

To be concrete, let us calculate the Casimir force between two parallel dielectric bodies in the vacuum.
We first consider the case of $N=5$ in which layers $1$, $3$ and $5$ are vacuum, $2$ and $4$ are dielectric bodies and $0$ and $6$ are perfect conductors.
Then we take the limit $\Delta z_1, \Delta z_5\to \infty$ to effectively make the effect of boundaries negligible.
In this limit we obtain $\tilRC{0}{3} \simeq \tilRC{1}{3}$ and $\tilRC{6}{3} \simeq \tilRC{5}{3}$ .
Noting that the reflection matrix is diagonal in the TM/TE basis, with use of Eqs.~(\ref{eq:recur_T}), (\ref{eq:recur_R}) and (\ref{eq:rel_jp1toj}), we obtain
\begin{align}
	\hat{R}_\lambda^{(1|3)} = \frac{ \hat{R}_\lambda^{(2|3)}+ \hat{R}_\lambda^{(2|1)}e^{-2k^{(2)}_z \Delta z_2}}
	{1- \hat{R}_\lambda^{(1|2)} \hat{R}_\lambda^{(3|2)}e^{-2k^{(2)}_z \Delta z_2}},~~~~~~
	\hat{R}_\lambda^{(5|3)} = \frac{ \hat{R}_\lambda^{(4|3)}+ \hat{R}_\lambda^{(4|5)}e^{-2k^{(4)}_z \Delta z_4}}
	{1- \hat{R}_\lambda^{(5|4)} \hat{R}_\lambda^{(3|4)}e^{-2k^{(4)}_z \Delta z_4}}.
\end{align}
Then, from Eq.~(\ref{eq:F_cas_j_formula_diagonal}), the Casimir force acting on the layer $3$ is given by
\begin{equation}
    F_\text{Cas}^{(0|3|6)} = - T \sum_{\lambda} {\sum_{\ell \geqslant 0}}' \int_{{\bm k}_\parallel}
    \frac{ 2 \hat{k}_{z, \ell\lambda}^{(3)}  e^{- 2 \hat{k}_{z,\ell\lambda}^{(3)} \Delta z_{3}} \,
    \hat{R}_\lambda^{(5|3)} \hat{R}_\lambda^{(1|3)}}{1 - e^{- 2\hat{k}_{z,\ell\lambda}^{(3)} \Delta z_{3}} \,\hat{R}^{(5|3)}_\lambda \hat{R}_\lambda^{(1|3)}},
\end{equation}
consistent with the standard formula~\cite{Bordag:2009zz}.
It is instructive to check again that this result is obtained from the Casimir energy (\ref{eq:Cas_energy_simple}):
\begin{align}
    \mathcal{F}_\text{Cas,r}^{(0|6)} \supset
    T {\sum_{\ell \geqslant 0}}' \int_{{\bm k}_\parallel}\ln \hat{f}^{(0|3|6)} (i \xi_\ell)
    =T \sum_{\lambda}{\sum_{\ell \geqslant 0}}' \int_{{\bm k}_\parallel}\ln \left(1-e^{-2 \hat k^{(3)}_{z,\ell\lambda}\Delta z_3}  \hat{R}_\lambda^{(5|3)} \hat{R}_\lambda^{(1|3)} \right),
\end{align}
where we have only considered the first term in Eq.~(\ref{eq:Cas_energy_simple}), since the other terms do not depend on $\Delta z_3$ and do not contribute to the Casimir force.

\section{Operational meaning}
\label{sec:phys_int}

\subsection{Casimir energy as work of system preparation}
\label{sec:prep}
Let us take a closer look at Eqs.~\eqref{eq:Cas_energy_simple}, \eqref{eq:F_Cas_j} and \eqref{eq:F_cas_j_formula} to have a better understanding of Casimir energy.
Consider two stacks of layers $(0|j)$ and $(j|N+1)$.
Suppose that we initially place $(0|j)$ and $(j|N+1)$ infinitely far away and bring them at a finite distance of $\Delta z_j$ by hand.
The amount of work required for this process is obtained from
\begin{align} 
    \mathcal{W}^{(0|j|N+1)} 
    &\equiv \int^{\Delta z_j}_\infty \dd \Delta z_j' \, \qty( - F_\text{Cas}^{(0|j|N+1)}  ) \\
    &= T {\sum_{\ell \geqslant 0}}' \int_{{\bm k}_\parallel} \ln \hat{f}^{(0|j|N+1)} (i \xi_\ell). \label{eq:W_j}
\end{align}
In the second line, we have used the fact that $\hat{f}^{(0|j|N+1)} (i \xi_\ell) \to 1$ for $\Delta z_j \to \infty$.
The resultant expression \eqref{eq:W_j} coincides with the first term in Eq.~\eqref{eq:Cas_energy_simple}.

Then what is the operational meaning of the remaining terms in Eq.~\eqref{eq:Cas_energy_simple}?
To clarify this, we first discuss $\ln \hat{f}^{(j|k|l)} ( i \xi_\ell )$ [see Eq.~\eqref{eq:char_general}].
Again consider two stacks of layer $(j|k)$ and $(k|l)$.
The Casimir force acting on the layer $k$ is obtained from
\begin{equation}
    F_\text{Cas}^{(j|k|l)} 
    \equiv
    - \qty( \frac{\partial \mathcal{F}_\text{Cas}^{(j|l)}}{\partial \Delta z_k} )_{\Delta z_{m \neq k}}
    = 
    - T {\sum_{\ell \geqslant 0}}' \int_{{\bm k}_{\parallel}} 
    \frac{\partial \ln \hat{f}^{(j|k|l)}( i \xi_\ell )}{\partial \Delta z_k}.
\end{equation}
Suppose that they are placed infinitely far away initially.
The amount of work required to bring them at a finite distance of $\Delta z_k$ is given by
\begin{equation}
    \mathcal{W}^{(j|k|l)} 
    \equiv 
    \int^{\Delta z_k}_\infty \dd \Delta z_k' \,
    \qty( - F_\text{Cas}^{(j|k|l)} ) 
    = 
    T {\sum_{\ell \geqslant 0}}' \int_{{\bm k}_{\parallel}} \ln \hat{f}^{(j|k|l)}( i \xi_\ell ).
\end{equation}
One can see that the case of Eq.~\eqref{eq:W_j} is a specific example of $(j|k|l) \to (0|j|N+1)$.

Now the meaning of each term in Eq.~\eqref{eq:Cas_energy_simple} is clear. 
As an illustration, we start with $\hat{f}^{(0|1|2)}(i \xi_\ell)$, which appears in the summation of the second term of Eq.~\eqref{eq:Cas_energy_simple}. 
This corresponds to the work required to bring $(0|1)$ and $(1|2)$ at a finite distance, namely
\begin{equation}
    \mathcal{W}^{(0|1|2)} = T {\sum_{\ell \geqslant 0}}' \int_{{\bm k}_{\parallel}} \ln \hat{f}^{(0|1|2)}( i \xi_\ell ),
\end{equation}
which implies that a body $(0|2)$ can be constructed with a cost of $\mathcal{W}^{(0|1|2)}$.
Then, add $(2|3)$ to the body $(0|2)$ from the right-hand side, \textit{i.e.}, a positive infinity $z = \infty$.
The amount of work for this process is
\begin{equation}
    \mathcal{W}^{(0|2|3)} = T {\sum_{\ell \geqslant 0}}' \int_{{\bm k}_{\parallel}} \ln \hat{f}^{(0|2|3)}( i \xi_\ell ).
\end{equation}
By repeating this procedure, we can construct a body $(0|j)$. 
The amount of work required for this whole process is
\begin{equation}
    \mathcal{W}^{(0|j)} 
    \equiv \sum_{k = 1}^{j - 1} \mathcal{W}^{(0|k|k+1)}
    = T {\sum_{\ell \geqslant 0}}' \int_{{\bm k}_{\parallel}} \sum_{k = 1}^{j - 1} \ln \hat{f}^{(0|k|k+1)}( i \xi_\ell ),
\end{equation}
which is nothing but the second term in Eq.~\eqref{eq:Cas_energy_simple}.
Similarly, one may readily show that the amount of energy to construct the body $(j|N+1)$, which is given as
\begin{equation}
    \mathcal{W}^{(j|N+1)} 
    \equiv \sum_{k=1}^{N-j} \mathcal{W}^{(N-k|N-k+1|N+1)}
    = T {\sum_{\ell \geqslant 0}}' \int_{{\bm k}_{\parallel}} \sum_{k = 1}^{N - j} \ln \hat{f}^{(N-k|N-k+1|N+1)}( i \xi_\ell ),
\end{equation}
coincides with the third term in Eq.~\eqref{eq:Cas_energy_simple}.
Therefore, the total amount of work to construct the whole body $(0|N+1)$ is obtained from the summation of the three contributions, which coincides with the particular expression of the Casimir energy in Eq.~\eqref{eq:Cas_energy_simple}:
\begin{eBox}
\begin{equation}
    \mathcal{W}^{(0|N+1)}
    = \mathcal{W}^{(0|j)} + \mathcal{W}^{(j|N+1)} + \mathcal{W}^{(0|j|N+1)} = \text{$\mathcal{F}_\text{Cas,r}^{(0|N+1)}$ given in Eq.~\eqref{eq:Cas_energy_simple}}.
\end{equation}
\end{eBox}

\subsection{Independence of construction and redundancies}
\label{sec:red}
In the previous section, we have shown that the each term in Eq.~\eqref{eq:Cas_energy_simple} corresponds to the amount of work in order to construct a body $(0|N+1)$ in the following way: 
(i) construct a body $(0|j)$ by adding a layer successively as $(0|1) + (1|2) \to (0|2)$, $(0|2) + (2|3) \to (0|3)$, $\cdots$, $(0|j-1) + (j-1|j) \to (0|j)$,
(ii) construct a body $(j|N+1)$ by adding a layer successively as $(N-1|N) + (N|N+1) \to (N-1|N+1)$, $\cdots$, $(j|j+1) + (j+1|N+1) \to (j|N+1)$,
(iii) finally obtain a body $(0|N+1)$ by merging $(0|j)$ and $(j|N+1)$.

Let us consider the same Casimir energy in a different expression.
For instance, the Casimir energy can be expressed by using $\hat{\tilde{f}}^{(0|k|N+1)}$ with $k \neq j$.
This expression suggests a different way of construction: (i') construct $(0|k)$, (ii') construct $(k|N+1)$ and then (iii') obtain $(0|N+1)$ by merging them.
As we have shown, the Casimir energy based on $\hat{\tilde{f}}^{(0|k|N+1)}$ with $k \neq j$ is the same as that of $\hat{\tilde{f}}^{(0|j|N+1)}$ because of Eq.~\eqref{eq:j-indep}, which implies that the total amount of cost never depends on how we construct the whole body $(0|N+1)$.
Furthermore, one may consider a more complicated way of construction, for instance, $(0|j)$ in the first step (i) can be obtained by merging $(0|k)$ and $(k|j)$ for $0 < k < j$, whose required work is
$
    \mathcal{W}^{(0|j)} = \mathcal{W}^{(0|k)} + \mathcal{W}^{(k|j)}
$.
Again the total amount of work to obtain $(0|j)$ never depends on how we construct $(0|j)$ because of Eq.~\eqref{eq:swap_k_l}.
In this way, a particular way of construction indicates a particular expression of the Casimir energy. 
The redundancy in expression of the multilayered Casimir energy simply means that there are many different ways of constructing the same body, but the total amount of work never depends on how we construct it.

Before closing this section, we remark that this redundancy is useful when we discuss the Casimir energy/force in a practical setup.
In a typical situation, some bodies are solid, whose thickness $\Delta z$ cannot be changed.
For instance, suppose that we have three solid bodies $(0|k)$, $(k|j)$, $(j|N+1)$ emersed in medium $j$ and $k$.
In this setup, the variable thickness parameters are $\Delta z_j$ and $\Delta z_k$ while the others are fixed.
One may express the Casimir energy in the following way
\begin{align} \label{eq:3bodies_1}
    \mathcal{F}^{(0|N+1)}_\text{Cas}
    &=
    \mathcal{W}^{(0|k)} + \mathcal{W}^{(k|j)} + \mathcal{W}^{(j|N+1)}
    + \mathcal{W}^{(k|j|N+1)} (\Delta z_j) + \mathcal{W}^{(0|k|N+1)} (\Delta z_j, \Delta z_k) \\
    &= 
    \mathcal{W}^{(0|k)} + \mathcal{W}^{(k|j)} + \mathcal{W}^{(j|N+1)}
    + \mathcal{W}^{(0|k|j)} (\Delta z_k) + \mathcal{W}^{(0|j|N+1)} (\Delta z_j, \Delta z_k). \label{eq:3bodies_2}
\end{align}
The first three terms are constant while only the last two terms depend on $\Delta z_j$ and $\Delta z_k$.
For this reason, this expression is useful since we can extract the terms that depend on $\Delta z_j$ and $\Delta z_k$.
Note that we have $\mathcal{W}^{(k|j|N+1)}+ \mathcal{W}^{(0|k|N+1)} = \mathcal{W}^{(0|k|j)} + \mathcal{W}^{(0|j|N+1)}$ because of Eq.~\eqref{eq:swap_k_l}.
The Casimir force acting on the $j$-side of the body $(k|j)$ can be obtained by taking a derivative with respect to $\Delta z_j$ with $\Delta z_k$ being fixed, namely
\begin{equation}
    F^{(0|j|N+1)}_\text{Cas}
    = - \qty( \frac{\partial \mathcal{W}^{(0|j|N+1)}}{\partial \Delta z_j}  )_{\Delta z_k}
    = - T {\sum_{\ell \geqslant 0}}' \int_{{\bm k}_\parallel} 
    \frac{\partial \ln  \hat{f}^{(0|j|N+1)} (i \xi_\ell)}{\partial \Delta z_j}.
\end{equation}
Owing to the redundancy of Eqs.~\eqref{eq:3bodies_1} and \eqref{eq:3bodies_2}, we also have
\begin{equation} \label{eq:j_force_red}
    F^{(0|j|N+1)}_\text{Cas}
    = - \qty( \frac{\partial \mathcal{W}^{(k|j|N+1)}}{\partial \Delta z_j}  )_{\Delta z_k}
    - \qty( \frac{\partial \mathcal{W}^{(0|k|N+1)}}{\partial \Delta z_j}  )_{\Delta z_k}.
\end{equation}
Similarly, the Casimir force acting on the $k$-side of the body is given by
\begin{equation}
    F^{(0|k|N+1)}_\text{Cas}
    = - \qty( \frac{\partial \mathcal{W}^{(0|k|N+1)}}{\partial \Delta z_k}  )_{\Delta z_j}
    = - \qty( \frac{\partial \mathcal{W}^{(0|k|j)}}{\partial \Delta z_k}  )_{\Delta z_j}
    - \qty( \frac{\partial \mathcal{W}^{(0|j|N+1)}}{\partial \Delta z_k}  )_{\Delta z_j}.
    \label{eq:k_force_red}
\end{equation}
As a consistency check, let us consider the case where the distance of the boundary bodies $(0|k)$ and $(j|N+1)$ are fixed while the body $(k|j)$ in between them can move.
The total Casimir force acting on the body $(k|j)$ is obtained by taking a derivative with respect to $\Delta z_j$ under the constraint $\Delta z_j + \Delta z_k = \text{const.}$, which reads
\begin{align}
    F^{(0|(k|j)|N+1)}_\text{Cas} 
    &=  - \qty( \frac{\partial \mathcal{F}_\text{Cas}^{(0|N+1)}}{\partial \Delta z_j}  )_{\Delta z_j + \Delta z_k}
    = - \qty( \frac{\partial \mathcal{W}^{(0|j|N+1)}}{\partial \Delta z_j}  )_{\Delta z_k}
    + \qty( \frac{\partial \mathcal{W}^{(0|j|N+1)}}{\partial \Delta z_k}  )_{\Delta z_j}
    + \qty( \frac{\partial \mathcal{W}^{(0|k|j)}}{\partial \Delta z_k}  )_{\Delta z_j} \\[.5em]
    &= 
    F^{(0|j|N+1)}_\text{Cas} - F^{(0|k|N+1)}_\text{Cas}.
\end{align}
Here we have used the chain rule and the redundancy \eqref{eq:k_force_red}.
The total Casimir force acting on the body $(k|j)$ is given by the difference between the force acting on the $j$-side and $k$-side as expected.

\section{Conclusions}
\label{sec:conc}

In this paper we have revisited the derivation of the Casimir energy and Casimir force for general multilayer setups.
First we have defined effective reflection/transmission matrices and the most general recursion relations for them.
These relations are essential ingredients for calculating the Casimir energy.
Second we have defined the characteristic function $f^{(i|k|j)}(\omega)$ so that the solution of $f^{(0|j|N+1)}(\omega)=0$ gives allowed mode in a given multilayer setup, and also carefully defined $\tilde f^{(i|k|j)}(\omega)$ in order to cancel unnecessary poles of $f^{(i|k|j)}(\omega)$.
With these preparations, we write down the explicit formula for the Casimir energy by applying the argument principle in complex analysis language. 
The Casimir energy contains UV divergence, and we need to subtract and renormalize it to obtain physically relevant results.
In order to extract UV dependence, we have carefully chosen the contour of complex integral and identified the origin of divergence.
We found that, the UV divergence is either properly subtracted by the counter term in the Lagrangian or does not contribute to the Casimir force in physically reasonable setups.
To the best of our knowledge, this is the first rigorous proof of the formula for the Casimir energy and force.
Although the resulting formula is essentially consistent with previous literature, our formulation makes treatments of UV dependence clear.
We also succeeded to give a clear interpretation of the Casimir energy as a work for preparing the system, which is originated from the proper treatment of poles.

\section*{Acknowledgment}
We would like to thank Y.~Ema for discussions at the early stage of this work.
This work was supported by World Premier International Research Center Initiative (WPI), MEXT, Japan.
This work was also supported by JSPS KAKENHI (Grant Numbers JP22K14044 [KM] and JP24K07010 [KN]).

\appendix
\section{Derivations}

Here we provide some explicit derivations omitted in the main text.

\subsection{Proof of Eq.~\eqref{eq:consistency_rel}}
\label{sec:proof_T}

The independence of the choice of the inserted layer, \textit{i.e.}, $\mathbb{T}^{(i|k|j)} = \mathbb{T}^{(i|l|j)}$ for $i < l < k < j$, can be shown if Eq.~\eqref{eq:consistency_rel} holds.
Here we provide the sketch of proof for Eq.~\eqref{eq:consistency_rel}:
\begin{align}
    &\qty[ e^{- i \mathbb{k}_z^{(k)}\Delta z_k}
    -
    \RC{j}{k}
    e^{i \mathbb{k}_z^{(k)}\Delta z_k}
    \RC{i}{k} ]
    \mathbb{T}^{(l|k)^{-1}}
    \qty[ e^{- i \mathbb{k}_z^{(l)}\Delta z_l}
    -
    \RC{k}{l}
    e^{i \mathbb{k}_z^{(l)}\Delta z_l}
    \RC{i}{l} ] \nonumber \\
    &=
    \qty[
            e^{- i \mathbb{k}_z^{(k)}\Delta z_k}
            -
            \RC{j}{k}
            e^{i \mathbb{k}_z^{(k)}\Delta z_k}
            \RC{l}{k}
        ]
    \mathbb{T}^{(l|k)^{-1}}
        \qty[
            e^{- i \mathbb{k}_z^{(l)}\Delta z_l}
            -
            \RC{j}{l}
            e^{i \mathbb{k}_z^{(l)}\Delta z_l}
            \RC{i}{l}
        ].
\end{align}
The left-hand side is given by
\begin{align}
    \text{(l.h.s.)} &= e^{- i \mathbb{k}_z^{(k)}\Delta z_k} \mathbb{T}^{(l|k)^{-1}} e^{- i \mathbb{k}_z^{(l)}\Delta z_l}
    -
    \RC{j}{k}
    e^{i \mathbb{k}_z^{(k)}\Delta z_k}
    \RC{i}{k}
    \mathbb{T}^{(l|k)^{-1}} e^{- i \mathbb{k}_z^{(l)}\Delta z_l}
    -
    e^{- i \mathbb{k}_z^{(k)}\Delta z_k} \mathbb{T}^{(l|k)^{-1}}
    \RC{k}{l}
    e^{i \mathbb{k}_z^{(l)}\Delta z_l}
    \RC{i}{l} \nonumber \\
    &\qquad 
    +
    \RC{j}{k}
    e^{i \mathbb{k}_z^{(k)}\Delta z_k}
    \RC{i}{k}
    \mathbb{T}^{(l|k)^{-1}}
    \RC{k}{l}
    e^{i \mathbb{k}_z^{(l)}\Delta z_l}
    \RC{i}{l},
\end{align}
while the right-hand side is
\begin{align}
    \text{(r.h.s.)}
    &=
    e^{- i \mathbb{k}_z^{(k)}\Delta z_k} \mathbb{T}^{(l|k)^{-1}}e^{- i \mathbb{k}_z^{(l)}\Delta z_l}
    -
    \RC{j}{k}
    e^{i \mathbb{k}_z^{(k)}\Delta z_k}
    \RC{l}{k}
    \mathbb{T}^{(l|k)^{-1}}e^{- i \mathbb{k}_z^{(l)}\Delta z_l}
    -
    e^{- i \mathbb{k}_z^{(k)}\Delta z_k} \mathbb{T}^{(l|k)^{-1}}
    \RC{j}{l}
    e^{i \mathbb{k}_z^{(l)}\Delta z_l}
    \RC{i}{l} \nonumber \\
    &\qquad +
    \RC{j}{k}
    e^{i \mathbb{k}_z^{(k)}\Delta z_k}
    \RC{l}{k}
    \mathbb{T}^{(l|k)^{-1}}
    \RC{j}{l}
    e^{i \mathbb{k}_z^{(l)}\Delta z_l}
    \RC{i}{l}.
\end{align}
Let us subtract the right-hand side from the left-hand side.
The first term cancels each other.
The second term involves the difference of $\RC{i}{k}$ and $\RC{l}{k}$, which can be expressed as
\begin{align}
    \RC{i}{k} - \RC{l}{k}
    &= \TC{k}{l} e^{i \mathbb{k}_z^{(l)}\Delta z_l} \RC{i}{l} \TCinv{i}{l} \TC{i}{k} \nonumber \\
    &= \TC{k}{l} e^{i \mathbb{k}_z^{(l)}\Delta z_l} \RC{i}{l} 
    \qty[
        e^{-i \mathbb{k}_z^{(l)}\Delta z_l} - \RC{k}{l} e^{i \mathbb{k}_z^{(l)}\Delta z_l} \RC{i}{l}
    ]^{-1} \TC{l}{k},
\end{align}
where we have used Eqs.~\eqref{eq:recur_R} and \eqref{eq:recur_T}.
By using this relation, the subtraction of the second term yields
\begin{equation}
    \qty[ \qty( \text{l.h.s.} ) - \qty( \text{r.h.s.} ) ]_\text{$2$nd} = -
    \RC{j}{k}
    e^{i \mathbb{k}_z^{(k)}\Delta z_k}
    \TC{k}{l} e^{i \mathbb{k}_z^{(l)}\Delta z_l} \RC{i}{l}
    \qty[
        e^{-i \mathbb{k}_z^{(l)}\Delta z_l} - \RC{k}{l} e^{i \mathbb{k}_z^{(l)}\Delta z_l} \RC{i}{l}
    ]^{-1} e^{- i \mathbb{k}_z^{(l)}\Delta z_l}.
    \label{eq:2-19-2nd}
\end{equation}
Similarly, the subtraction of the third term gives
\begin{equation}
    \qty[ \qty( \text{l.h.s.} ) - \qty( \text{r.h.s.} ) ]_\text{$3$rd} =
    \RC{j}{k}
    \qty[
        e^{-i \mathbb{k}_z^{(k)}\Delta z_k} - \RC{l}{k} e^{i \mathbb{k}_z^{(k)}\Delta z_k} \RC{j}{k}
    ]^{-1} \TC{k}{l} e^{i \mathbb{k}_z^{(l)}\Delta z_l}
    \RC{i}{l}
    \label{eq:2-19-3rd}
\end{equation}
The subtraction of the fourth term gives
\begin{align}
    \qty[ \qty( \text{l.h.s.} ) - \qty( \text{r.h.s.} ) ]_\text{$4$th} &=
    \RC{j}{k}
    e^{i \mathbb{k}_z^{(k)}\Delta z_k}
    \qty{
        \RC{l}{k}
        +
        \TC{k}{l} e^{i \mathbb{k}_z^{(l)}\Delta z_l} \RC{i}{l} 
    \qty[
        e^{-i \mathbb{k}_z^{(l)}\Delta z_l} - \RC{k}{l} e^{i \mathbb{k}_z^{(l)}\Delta z_l} \RC{i}{l}
    ]^{-1} \TC{l}{k}
    } \nonumber \\
    &\qquad \times \mathbb{T}^{(l|k)^{-1}}
    \RC{k}{l}
    e^{i \mathbb{k}_z^{(l)}\Delta z_l}
    \RC{i}{l} \nonumber \\
    & -
    \RC{j}{k}
    e^{i \mathbb{k}_z^{(k)}\Delta z_k}
    \RC{l}{k}
    \mathbb{T}^{(l|k)^{-1}} \nonumber \\
    &\qquad \times \qty{
        \RC{k}{l}
        +
        \TC{l}{k} e^{i \mathbb{k}_z^{(k)}\Delta z_k} \RC{j}{k} \qty[
            e^{-i \mathbb{k}_z^{(k)}\Delta z_k} - \RC{l}{k} e^{i \mathbb{k}_z^{(k)}\Delta z_k} \RC{j}{k}
        ]^{-1} \TC{k}{l}
    }
    e^{i \mathbb{k}_z^{(l)}\Delta z_l}
    \RC{i}{l} \nonumber \\
    & =
    \RC{j}{k}
    e^{i \mathbb{k}_z^{(k)}\Delta z_k}
    \TC{k}{l} e^{i \mathbb{k}_z^{(l)}\Delta z_l} \RC{i}{l} 
    \qty[
        e^{-i \mathbb{k}_z^{(l)}\Delta z_l} - \RC{k}{l} e^{i \mathbb{k}_z^{(l)}\Delta z_l} \RC{i}{l}
    ]^{-1} \RC{k}{l}
    e^{i \mathbb{k}_z^{(l)}\Delta z_l}
    \RC{i}{l} \label{eq:2-19-4th-1} \\
    & -
    \RC{j}{k}
    e^{i \mathbb{k}_z^{(k)}\Delta z_k}
    \RC{l}{k} e^{i \mathbb{k}_z^{(k)}\Delta z_k} \RC{j}{k} \qty[
        e^{-i \mathbb{k}_z^{(k)}\Delta z_k} - \RC{l}{k} e^{i \mathbb{k}_z^{(k)}\Delta z_k} \RC{j}{k}
    ]^{-1} \TC{k}{l} e^{i \mathbb{k}_z^{(l)}\Delta z_l}
    \RC{i}{l}.
    \label{eq:2-19-4th-2}
\end{align}
Again we have used Eqs.~\eqref{eq:recur_R} and \eqref{eq:recur_T}.
Summing up Eqs.~\eqref{eq:2-19-2nd} and \eqref{eq:2-19-4th-1}, we find
\begin{align}
    \text{\eqref{eq:2-19-2nd}} - \text{\eqref{eq:2-19-4th-1}}
    &=
    - \RC{j}{k}
    e^{i \mathbb{k}_z^{(k)}\Delta z_k}
    \TC{k}{l} e^{i \mathbb{k}_z^{(l)}\Delta z_l} \RC{i}{l}
    \qty[
        e^{-i \mathbb{k}_z^{(l)}\Delta z_l} - \RC{k}{l} e^{i \mathbb{k}_z^{(l)}\Delta z_l} \RC{i}{l}
    ]^{-1} \qty[e^{- i \mathbb{k}_z^{(l)}\Delta z_l} -  \RC{k}{l}
    e^{i \mathbb{k}_z^{(l)}\Delta z_l}\RC{i}{l}] \nonumber \\
    &=
    - \RC{j}{k}
    e^{i \mathbb{k}_z^{(k)}\Delta z_k}
    \TC{k}{l} e^{i \mathbb{k}_z^{(l)}\Delta z_l} \RC{i}{l}.
    \label{eq:2-19-1}
\end{align}
Similarly, the sum of Eqs.~\eqref{eq:2-19-3rd} and \eqref{eq:2-19-4th-2} gives
\begin{align}
    \text{\eqref{eq:2-19-3rd}} - \text{\eqref{eq:2-19-4th-2}}
    &=
    \RC{j}{k} e^{i \mathbb{k}_z^{(k)}\Delta z_k} \qty[ e^{-i \mathbb{k}_z^{(k)}\Delta z_k} - \RC{l}{k} e^{i \mathbb{k}_z^{(k)}\Delta z_k} \RC{j}{k} ]
    \qty[
        e^{-i \mathbb{k}_z^{(k)}\Delta z_k} - \RC{l}{k} e^{i \mathbb{k}_z^{(k)}\Delta z_k} \RC{j}{k}
    ]^{-1} \TC{k}{l} e^{i \mathbb{k}_z^{(l)}\Delta z_l} \RC{i}{l} \nonumber \\
    &= \RC{j}{k} e^{i \mathbb{k}_z^{(k)}\Delta z_k}\TC{k}{l} e^{i \mathbb{k}_z^{(l)}\Delta z_l} \RC{i}{l}.
    \label{eq:2-19-2}
\end{align}
Now it is clear that Eqs.~\eqref{eq:2-19-1} and \eqref{eq:2-19-2} cancel each other.
Therefore, the subtraction of the right-hand side from the left-hand side gives zero, which completes the proof of Eq.~\eqref{eq:consistency_rel}.

\subsection{Proof of $\mathbb{R}^{(i|k|j)} = \mathbb{R}^{(i|l|j)}$}
\label{sec:proof_R}

Here we briefly sketch the proof for the reflection coefficient corresponding to that for the transmission coefficient given in Sec.~\ref{sec:proof_T}.
By using the recursion relations \eqref{eq:recur_T} and \eqref{eq:recur_R}, we can express the reflection coefficient $\mathbb{R}^{(i|j)}$ in two ways:
\begin{align}
    \mathbb{R}^{(i|k|j)} &=
    \RC{k}{j} + \TC{j}{k} e^{i \mathbb{k}_z^{(k)}\Delta z_k} \RC{i}{k} \TCinv{i}{k} \TC{i}{j} \nonumber
    \\
    &=
    \RC{k}{j} +
    \TC{j}{k} e^{i \mathbb{k}_z^{(k)}\Delta z_k} \RC{l}{k} \TCinv{l}{k}\qty[  e^{-i \mathbb{k}_z^{(l)}\Delta z_l} - \RC{k}{l} e^{i \mathbb{k}_z^{(l)}\Delta z_l} \RC{i}{l} ]\TCinv{i}{l} \TC{i}{j} \nonumber \\
    &\qquad +
    \TC{j}{k} e^{i \mathbb{k}_z^{(k)}\Delta z_k} \TC{k}{l} e^{i \mathbb{k}_z^{(l)}\Delta z_l} \RC{i}{l} \TCinv{i}{l} \TC{i}{j}, \label{eq:Rij_1}
\end{align}
and
\begin{align}
    \mathbb{R}^{(i|l|j)} &=
    \RC{l}{j} + \TC{j}{l} e^{i \mathbb{k}_z^{(l)}\Delta z_l} \RC{i}{l} \TCinv{i}{l} \TC{i}{j} \nonumber
    \\
    &=
    \RC{k}{j} + 
    \TC{j}{k} e^{i \mathbb{k}_z^{(k)}\Delta z_k} \RC{l}{k} \TCinv{l}{k}\qty[  e^{-i \mathbb{k}_z^{(l)}\Delta z_l} - \RC{k}{l} e^{i \mathbb{k}_z^{(l)}\Delta z_l} \RC{i}{l} ]\TCinv{i}{l} \TC{i}{j} \nonumber \\
    & \qquad -
    \TC{j}{k} e^{i \mathbb{k}_z^{(k)}\Delta z_k} \RC{l}{k} e^{i \mathbb{k}_z^{(l)}\Delta z_l} \RC{j}{k}\qty[
        e^{-i \mathbb{k}_z^{(k)}\Delta z_k} - \RC{l}{k} e^{i \mathbb{k}_z^{(k)}\Delta z_k} \RC{j}{k}
    ]^{-1} \TC{k}{l} e^{i \mathbb{k}_z^{(l)}\Delta z_l} \RC{i}{l} \TCinv{i}{l} \TC{i}{j} \nonumber
    \\
    & \qquad +
    \TC{j}{k} \qty[ e^{-i \mathbb{k}_z^{(k)}\Delta z_k} - \RC{l}{k} e^{i \mathbb{k}_z^{(k)}\Delta z_k} \RC{j}{k} ]^{-1} \TC{k}{l} e^{i \mathbb{k}_z^{(l)}\Delta z_l} \RC{i}{l} \TCinv{i}{l} \TC{i}{j}. \label{eq:Rij_2}
\end{align}
The first terms in Eq.~\eqref{eq:Rij_1} and Eq.~\eqref{eq:Rij_2} cancel each other.
Noticing that the second and third terms of Eq.~\eqref{eq:Rij_2} can be summarized as
\begin{align}
    \qty[\text{($2$nd)} - \text{($3$rd)}]_\text{Eq.~\eqref{eq:Rij_2}} &=
    \TC{j}{k} e^{i \mathbb{k}_z^{(k)}\Delta z_k}
    \qty[
        e^{-i \mathbb{k}_z^{(k)}\Delta z_k} - \RC{l}{k} e^{i \mathbb{k}_z^{(k)}\Delta z_k} \RC{j}{k}
    ] \qty[
        e^{-i \mathbb{k}_z^{(k)}\Delta z_k} - \RC{l}{k} e^{i \mathbb{k}_z^{(k)}\Delta z_k} \RC{j}{k}
    ]^{-1}\nonumber\\
    & \qquad \times \TC{k}{l} e^{i \mathbb{k}_z^{(l)}\Delta z_l} \RC{i}{l} \TCinv{i}{l} \TC{i}{j} \nonumber \\
    &= \TC{j}{k} e^{i \mathbb{k}_z^{(k)}\Delta z_k}
    \TC{k}{l} e^{i \mathbb{k}_z^{(l)}\Delta z_l} \RC{i}{l} \TCinv{i}{l} \TC{i}{j},
\end{align}
one concludes that the second term of Eq.~\eqref{eq:Rij_1} is also canceled out by the second and third terms of Eq.~\eqref{eq:Rij_2}.
This completes the proof of $\mathbb{R}^{(i|k|j)} = \mathbb{R}^{(i|l|j)}$.

\subsection{Proof of Eq.~\eqref{eq:j-indep}}
\label{sec:proof_j-indep}

Starting from the definition given in Eq.~\eqref{eq:argument_principle}, one may immediately find
\begin{align}
    \tilde f^{(0|j|N+1)}(\omega) &= f^{(0|j|N+1)}(\omega)
    \qty[ \prod_{k=1}^{j-1} f^{(0|j-k|j-k+1)} (\omega) ]
    \qty[ \prod_{k=1}^{N-j} f^{(j+k-1|j+k|N+1)} (\omega) ] \nonumber \\
    &=
    f^{(0|j|N+1)}(\omega)
    \qty[ \prod_{k=1}^{j-1} f^{(0|j-k|j-k+1)} (\omega) ]
    \qty[ \prod_{k=1}^{N-j-1} f^{(j+k-1|j+k|N+1)} (\omega) ]
    f^{(j|j+1|N+1)} (\omega) \nonumber \\
    &=
    f^{(0|j+1|N+1)}(\omega)
    f^{(0|j|j+1)} (\omega)
    \qty[ \prod_{k=1}^{j-1} f^{(0|j-k|j-k+1)} (\omega) ]
    \qty[ \prod_{k=1}^{N-j-1} f^{(j+k-1|j+k|N+1)} (\omega) ]\nonumber \\
    &=
    f^{(0|j+1|N+1)}(\omega)
    \qty[ \prod_{k=1}^{j} f^{(0|j+1-k|j-k+2)} (\omega) ]
    \qty[ \prod_{k=1}^{N-j-1} f^{(j+k-1|j+k|N+1)} (\omega) ]
    = \tilde f^{(0|j+1|N+1)}(\omega).
\end{align}
Here we have used Eq.~\eqref{eq:swap_k_l}, \textit{i.e.}, $   f^{(0|j+1|N+1)}(\omega)f^{(0|j|j+1)} (\omega) = f^{(0|j|N+1)}(\omega) f^{(j|j+1|N+1)} (\omega)$.

\subsection{Proof of Eqs.~\eqref{eq:R0j_new} and \eqref{eq:hatRC0j}}
Let us rewrite Eq.~\eqref{eq:recur_R} as
\begin{align}
    \RC{0}{j} &= \qty[\RC{j-1}{j} \TCinv{j-1}{j} - \qty( \RC{j-1}{j} \TCinv{j-1}{j} \RC{j}{j-1} - \TC{j}{j-1} )e^{i \mathbb{k}_z^{(j-1)}\Delta z_{j-1}} \RC{0}{j-1}  e^{i \mathbb{k}_z^{(j-1)}\Delta z_{j-1}}
    ] \nonumber \\
    & \qquad \times \qty[
        \TCinv{j-1}{j} -\TCinv{j-1}{j} \RC{j}{j-1} e^{i \mathbb{k}_z^{(j-1)}\Delta z_{j-1}} \RC{0}{j-1}  e^{i \mathbb{k}_z^{(j-1)}\Delta z_{j-1}}
    ]^{-1}.
\end{align}
By using Eq.~\eqref{eq:rel_jm1toj}, we obtain Eq.\eqref{eq:R0j_new} as
\begin{align}
    \RC{0}{j} &= \qty[- \TCinv{j-1}{j}\RC{j}{j-1} e^{- i \mathbb{k}_z^{(j-1)}\Delta z_{j-1}} + \TCinv{j-1}{j}
    e^{i \mathbb{k}_z^{(j-1)}\Delta z_{j-1}} \RC{0}{j-1} 
    ] \nonumber \\
    & \qquad \times \qty[
        \TCinv{j-1}{j} e^{- i \mathbb{k}_z^{(j-1)}\Delta z_{j-1}} -\TCinv{j-1}{j} \RC{j}{j-1} e^{i \mathbb{k}_z^{(j-1)}\Delta z_{j-1}} \RC{0}{j-1}
    ]^{-1}.
\end{align}
By using Eqs.~\eqref{eq:chr_highE}, ~\eqref{eq:analytic_c_R} and \eqref{eq:analytic_c_T}, we obtain its expression on the contour $\mathcal{C}_{R_1+I_1}$ as
\begin{align}
    \tilRC{0}{j} &= \qty[- \tilTCinv{j-1}{j} \tilRC{j}{j-1} e^{\hat{\mathbb{k}}_z^{(j-1)}\Delta z_{j-1}}  + \tilTCinv{j-1}{j}
    e^{- \hat{\mathbb{k}}_z^{(j-1)}\Delta z_{j-1}} \tilRC{0}{j-1}
    ] \nonumber \\
    & \qquad \times \qty[
        \tilTCinv{j-1}{j} e^{\hat{\mathbb{k}}_z^{(j-1)}\Delta z_{j-1}} -\tilTCinv{j-1}{j} \tilRC{j}{j-1} e^{-\hat{\mathbb{k}}_z^{(j-1)}\Delta z_{j-1}} \tilRC{0}{j-1}  
    ]^{-1}.
    \label{eq:hatRC0japp}
\end{align}
Now we can also express $\hatRC{0}{j}$ as
\begin{align}
    \hatRC{0}{j} &= \qty[- \hatTCinv{j-1}{j} \hatRC{j}{j-1} e^{ \check{\mathbb{k}}_z^{(j-1)}\Delta z_{j-1}}  + \hatTCinv{j-1}{j}
    e^{- \check{\mathbb{k}}_z^{(j-1)}\Delta z_{j-1}} \hatRC{0}{j-1}
    ] \nonumber \\
    & \qquad \times \qty[
        \hatTCinv{j-1}{j} e^{\check{\mathbb{k}}_z^{(j-1)}\Delta z_{j-1}} -\hatTCinv{j-1}{j} \hatRC{j}{j-1} e^{-\check{\mathbb{k}}_z^{(j-1)}\Delta z_{j-1}} \hatRC{0}{j-1}
    ]^{-1} \nonumber \\
    &=
    \qty[- \tilTCinv{j-1}{j}_P \tilRC{j}{j-1}_P e^{- \hat{\mathbb{k}}_{z P}^{(j-1)} \Delta z_{j-1}}
    + \tilTCinv{j-1}{j}_P e^{- \hat{\mathbb{k}}_{z P}^{(j-1)} \Delta z_{j-1}} \tilRCinv{0}{j-1}_P
    ] \nonumber \\
    & \qquad \times \qty[
        \tilTCinv{j-1}{j}_P e^{- \hat{\mathbb{k}}_{z P}^{(j-1)} \Delta z_{j-1}} - \tilTCinv{j-1}{j}_P \tilRC{j}{j-1} e^{\hat{\mathbb{k}}_{z P}^{(j-1)} \Delta z_{j-1}} \tilRCinv{0}{j-1}_P
    ]^{-1}
    = \tilRCinv{0}{j}_P,
\end{align}
which coincides with Eq.~\eqref{eq:hatRC0j}.
Here we have used $\hatRC{0}{j-1} = \tilRCinv{0}{j-1}_P$.

\subsection{From Eq.~\eqref{eq:chr_recur1} to Eq.~\eqref{eq:chr_recur2}}
\label{sec:proof_chr_recur}

Let us start from Eq.~\eqref{eq:chr_recur1}.
Inserting Eqs.~\eqref{eq:char_general} and \eqref{eq:hatRC0japp} into Eq.~\eqref{eq:chr_recur1}, we obtain
\begin{align}
    &\det \Bigg[ \tilRCinv{0}{k} \Bigg]
    \det \Bigg[ \mathbb{1} - e^{\hat{\mathbb{k}}_z^{(k-1)}\Delta z_{k-1}} \tilRC{k}{k-1} e^{\hat{\mathbb{k}}_z^{(k-1)}\Delta z_{k-1}} \tilRCinv{0}{k-1} \Bigg] \nonumber \\
    &=
    \det \qty[- \tilRC{k}{k-1} e^{ \hat{\mathbb{k}}_z^{(k-1)}\Delta z_{k-1}} +
    e^{- \hat{\mathbb{k}}_z^{(k-1)}\Delta z_{k-1}} \tilRC{0}{k-1}
    ]^{-1} \det \qty[ e^{ \hat{\mathbb{k}}_z^{(k-1)}\Delta z_{k-1}} -\RC{k}{k-1} e^{-\hat{\mathbb{k}}_z^{(k-1)}\Delta z_{k-1}} \RC{0}{k-1}
    ] \nonumber \\
    & \qquad \times \det \qty[ \mathbb{1} - e^{\hat{\mathbb{k}}_z^{(k-1)}\Delta z_{k-1}} \tilRC{k}{k-1} e^{\hat{\mathbb{k}}_z^{(k-1)}\Delta z_{k-1}} \tilRCinv{0}{k-1} ] \nonumber \\
    &=
    \det \qty[ \mathbb{1} - e^{- \hat{\mathbb{k}}_z^{(k-1)} \Delta z_{k-1} } \tilRC{k}{k-1} e^{- \hat{\mathbb{k}}_z^{(k-1)} \Delta z_{k-1} } \tilRC{0}{k-1}  ] \det \Bigg[ e^{2 \hat{\mathbb{k}}_z^{(k-1)} \Delta z_{k-1} } \Bigg] \det \Bigg[ \tilRCinv{0}{k-1} \Bigg] \nonumber \\
    &=
    \hat{f}^{(0|k-1|k)} (\omega) \det \Bigg[ e^{2 \hat{\mathbb{k}}_z^{(k-1)} \Delta z_{k-1} } \Bigg] \det \Bigg[ \tilRCinv{0}{k-1} \Bigg],
\end{align}
which completes the proof of Eq.~\eqref{eq:chr_recur2}.

\section{Examples of reflection and transmission coefficients}
\label{app:reflection}

In this Appendix we summarize derivations for reflection and transmission coefficients for adjacent layers, which are basic ingredients for the general multilayer system.

\subsection{Dielectric material}
\label{app:dielectric}

\subsubsection*{Boundary conditions}

\begin{figure}[t]
\begin{center}
    \includegraphics[width=16cm]{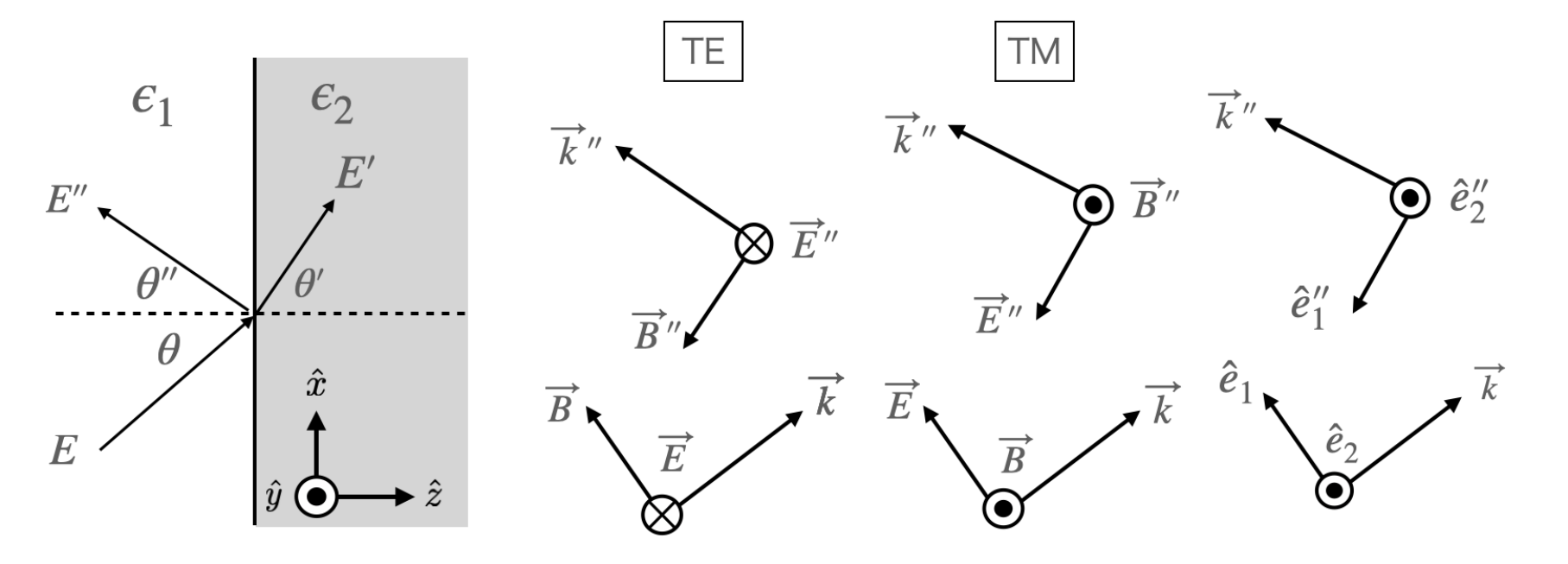}
    \end{center}
    \caption{(Left) Incoming photon, refracted photon and reflected photon with the strength of the electric field $E$, $E'$ and $E''$, respectively. (Right) Configuration of electric/magnetic fields for the TE and TM modes.}
    \label{fig:boundary}
\end{figure}

Let us consider electromagnetic waves incoming from layer 1 toward layer 2 as in Fig.~\ref{fig:boundary}.
Obviously we can think of layer 1 and 2 as $j$ and $j+1$ respectively, but we use the former convention for notational simplicity.
The boundary conditions of the electric/magnetic field at the boundary of the material with dielectric constant $\epsilon_i$ and magnetic permeability $\mu_i$ $(i=1,2)$ are given by
\begin{align}
	&\vec E_{1\parallel} = \vec E_{2\parallel},~~~~~~\epsilon_1E_{1\perp} - \epsilon_2E_{2\perp} = \sigma, \label{boundary_E}\\
	&B_{1\perp} = B_{2\perp},~~~~~~\frac{\vec B_{1\parallel}}{\mu_1} - \frac{\vec B_{2\parallel}}{\mu_2} = \vec K_{\parallel} \label{boundary_B},
\end{align}
where $\parallel$ and $\perp$ represent the direction parallel and perpendicular to the surface, $\sigma$ is the surface charge density and $K_\parallel$ is the current density on the surface.
In the following, we consider the dielectric material in which $\sigma=0$ and $\vec K_\parallel = 0$.

Generally incoming, transmitted and reflected electric/magnetic fields are written as
\begin{align}
	&\vec E(\vec x,t) = \vec{\mathcal E}e^{i(\vec k\cdot\vec x-\omega t)},~~~~~~
	\vec B(\vec x,t) = \vec{\mathcal B}e^{i(\vec k\cdot\vec x-\omega t)},\\
	&\vec E'(\vec x,t) = \vec{\mathcal E}' e^{i(\vec k'\cdot\vec x-\omega t)},~~~~~~
	\vec B'(\vec x,t) = \vec{\mathcal B}'e^{i(\vec k'\cdot\vec x-\omega t)},\\
	&\vec E''(\vec x,t) = \vec{\mathcal E}''e^{i(\vec k''\cdot\vec x-\omega t)},~~~~~~
	\vec B''(\vec x,t) = \vec{\mathcal B}''e^{i(\vec k''\cdot\vec x-\omega t)},
\end{align}
where $k=k''=\sqrt{\epsilon_1\mu_1}\omega$ and $k'=\sqrt{\epsilon_2\mu_2}\omega$ and $\vec B = \vec k\times \vec E/\omega$ and so on. 
Note that $\vec E_1= \vec E + \vec E'', \vec E_2= \vec E'$ and so on.
Without loss of generality, we take $z=0$ as a surface of the boundary.
The boundary condition at $z=0$ with any $(t,\vec x_\parallel)$ immediately implies common $\omega$ and $\vec k_\parallel = \vec k'_\parallel = \vec k''_\parallel$, which means $k''_{z} = -k_z$.
Thus we have $\theta''=\theta$ and 
\begin{align}
	k \sin \theta = k' \sin \theta'~~~~~~~\rightarrow~~~~~~\sin\theta' = \sqrt{\frac{\mu_1\epsilon_1}{\mu_2\epsilon_2}}\sin\theta.
\end{align}
This implies that, if $\mu_2\epsilon_2 \leq \mu_1\epsilon_1$, there is a threshold angle $\theta_{\rm ref}$ above which the incoming wave is totally reflected, i.e., $\theta' \geq \pi/2$. It is given by $\sin \theta_{\rm ref} = \sqrt{\frac{\mu_2\epsilon_2}{\mu_1\epsilon_1}}$.
Below we drop common factor $e^{i(\vec k_\parallel\cdot \vec x_\parallel -\omega t)}$.

\subsubsection*{Reflection coefficient for TM/TE basis}

There are two basic modes of the electromagnetic wave, TE and TM modes (see Fig.~\ref{fig:boundary}). Supposing that the surface is located at $z=0$, the TE (TM) mode has no electric (magnetic) component along the $z$ direction.\footnote{
    Here the terminology TM (transverse magnetic) and TE (transverse electric) refer to transverse with respect to $z$ axis, not to the wave vector.
}
By solving the boundary condition (\ref{boundary_E}) and (\ref{boundary_B}), we obtain the reflection coefficient.
For the TE mode, we obtain~\cite{Jackson:1998nia}
\begin{align}
	R_{\rm TE}^{(2,1)}(\theta) = \frac{\mathcal E''}{\mathcal E} 
	 = \frac{\mu_2 k_z - \mu_1 k_z'}{\mu_2 k_z + \mu_1 k_z'}
	 = \frac{\sqrt{\mu_1\epsilon_1}\cos\theta -\frac{\mu_1}{\mu_2} \sqrt{\mu_2\epsilon_2-\mu_1\epsilon_1\sin^2\theta}} 
	{\sqrt{\mu_1\epsilon_1}\cos\theta +\frac{\mu_1}{\mu_2} \sqrt{\mu_2\epsilon_2-\mu_1\epsilon_1\sin^2\theta}},
	\label{rTE}
\end{align}
and
\begin{align}
	T_{\rm TE}^{(2,1)}(\theta) = \frac{\mathcal E'}{\mathcal E} 
	= \frac{2\mu_2 k_z }{\mu_2 k_z + \mu_1 k_z'}.
	\label{tTE}
\end{align}
For the TM mode, we obtain
\begin{align}
	R_{\rm TM}^{(2,1)} (\theta) = \frac{\mathcal E''}{\mathcal E} 
	= \frac{\epsilon_2 k_z - \epsilon_1 k_z'}{\epsilon_2 k_z + \epsilon_1 k_z'} 
	= \frac{\frac{\mu_1}{\mu_2}\mu_2\epsilon_2\cos\theta -\sqrt{\mu_1\epsilon_1} \sqrt{\mu_2\epsilon_2-\mu_1\epsilon_1\sin^2\theta}}
	{\frac{\mu_1}{\mu_2}\mu_2\epsilon_2\cos\theta +\sqrt{\mu_1\epsilon_1} \sqrt{\mu_2\epsilon_2-\mu_1\epsilon_1\sin^2\theta}}.
	\label{rTM}
\end{align}
and
\begin{align}
	T_{\rm TM}^{(2,1)}(\theta) = \frac{\mathcal E'}{\mathcal E} 
	= \sqrt{\frac{\epsilon_1\epsilon_2\mu_2}{\mu_1}}\frac{2 k_z }{\epsilon_2 k_z + \epsilon_1 k_z'}.
	\label{tTM}
\end{align}
Thus there is no mixing between TM and TE modes. Usually we consider the case of $\mu_1=\mu_2=1$.
These coefficients for the inverse propagation direction is obviously obtained by exchanging $1\leftrightarrow 2$, $k_z \leftrightarrow k_z'$. 
One can easily check the relation (\ref{eq:rel_jp1toj}) such as $\left(R_{\lambda}^{(2,1)}\right)^2 + T_{\lambda}^{(2,1)}T_{\lambda}^{(1,2)} = 1$ for $\lambda = {\rm TM, TE}$.

\subsubsection*{Reflection coefficient for right/left circular basis}

Let us define the right/left circular polarization basis as
\begin{align}
	\hat e_{R,L} = \frac{1}{\sqrt 2}(\hat e_1 \pm i\hat e_2)
	=\frac{1}{\sqrt 2 k}\begin{pmatrix} k_z \\ \pm ik \\ -k_x \end{pmatrix},~~~~~~~~~~~~
	\hat e''_{R,L} = \frac{1}{\sqrt 2}(\hat e''_1 \pm i\hat e''_2)
	=\frac{1}{\sqrt 2 k}\begin{pmatrix} -k_z \\ \pm ik \\ -k_x \end{pmatrix},
\end{align}
where $\hat e_1$ and $\hat e_2$ are defined with respect to the wave vector as shown in Fig.~\ref{fig:boundary}.
We write the electric field as
\begin{align}
	&\vec E = e^{ik_z z}(\mathcal E_R \hat e_R + \mathcal E_L \hat e_L),\\
	&\vec E' = e^{ik'_z z}(\mathcal E'_R \hat e'_R + \mathcal E'_L \hat e'_L),\\
	&\vec E'' = e^{ik''_z z}(\mathcal E''_R \hat e''_R + \mathcal E''_L \hat e''_L).
\end{align}
By noting $\vec B = \vec k \times \vec E/\omega$, the magnetic field is written as
\begin{align}
	&\vec B = e^{ik_z z}\sqrt{\epsilon_1}(\mathcal E_R \hat e^{(B)}_R + \mathcal E_L \hat e^{(B)}_L),\\
	&\vec B' = e^{ik'_z z}\sqrt{\epsilon_2}(\mathcal E'_R \hat e^{'(B)}_R + \mathcal E'_L \hat e^{'(B)}_L),\\
	&\vec B'' = e^{ik''_z z}\sqrt{\epsilon_1}(\mathcal E''_R \hat e^{''(B)}_R + \mathcal E''_L \hat e^{''(B)}_L).
\end{align}
where
\begin{align}
	\hat e^{(B)}_{R,L} 
	=\frac{1}{\sqrt 2 k}\begin{pmatrix} \mp ik_z \\ k \\ \pm ik_x \end{pmatrix},~~~~~~
	\hat e^{'(B)}_{R,L} 
	=\frac{1}{\sqrt 2 k'}\begin{pmatrix} \mp ik'_z \\ k' \\ \pm ik'_x \end{pmatrix},~~~~~~
	\hat e^{''(B)}_{R,L} 
	=\frac{1}{\sqrt 2 k}\begin{pmatrix} \pm ik_z \\  k \\ \pm ik_x \end{pmatrix}.
\end{align}
with $k_x'=k_x$. 

First we consider the incoming right-handed wave: $\mathcal E_L=0$.
From the boundary conditions (\ref{boundary_E}) and (\ref{boundary_B}), we obtain\footnote{
	Among six boundary conditions, four of them are independent. We can choose, for example, 
	$E_x, E_y, E_z$ and $B_z$.
}
\begin{align}
	&R^{(2,1)}_{RR} \equiv \frac{\mathcal E''_R}{\mathcal E_R}= \frac{1}{2}\left( \frac{\epsilon_2 k_z - \epsilon_1 k_z'}{\epsilon_2 k_z + \epsilon_1 k_z'} 
	 + \frac{k_z-k_z'}{k_z + k_z'}\right) = \frac{1}{2}\left( R_{\rm TM}^{(2,1)} + R_{\rm TE}^{(2,1)} \right)\\
	&R^{(2,1)}_{LR}\equiv \frac{\mathcal E''_L}{\mathcal E_R}= \frac{1}{2}\left( \frac{\epsilon_2 k_z - \epsilon_1 k_z'}{\epsilon_2 k_z + \epsilon_1 k_z'} 
	 - \frac{k_z-k_z'}{k_z + k_z'}\right) = \frac{1}{2}\left( R_{\rm TM}^{(2,1)} - R_{\rm TE}^{(2,1)} \right).
\end{align}
Similarly, for incoming left-handed wave, $\mathcal E_R=0$, we obtain
\begin{align}
	&R^{(2,1)}_{RL} \equiv \frac{\mathcal E''_R}{\mathcal E_L}= \frac{1}{2}\left( \frac{\epsilon_2 k_z - \epsilon_1 k_z'}{\epsilon_2 k_z + \epsilon_1 k_z'} 
	 - \frac{k_z-k_z'}{k_z + k_z'}\right) = \frac{1}{2}\left( R_{\rm TM}^{(2,1)} - R_{\rm TE}^{(2,1)} \right)\\
	&R^{(2,1)}_{LL}\equiv \frac{\mathcal E''_L}{\mathcal E_L}= \frac{1}{2}\left( \frac{\epsilon_2 k_z - \epsilon_1 k_z'}{\epsilon_2 k_z + \epsilon_1 k_z'} 
	 + \frac{k_z-k_z'}{k_z + k_z'}\right) = \frac{1}{2}\left( R_{\rm TM}^{(2,1)} + R_{\rm TE}^{(2,1)} \right).
\end{align}
Transmission coefficients are also obtained in a similar way.
As a result, the reflection matrix is given by
\begin{align}	
	\begin{pmatrix} \mathcal E_L'' \\ \mathcal E_R''  \end{pmatrix}
	= \RC{2}{1}
	\begin{pmatrix} \mathcal E_R \\ \mathcal E_L \end{pmatrix},
	~~~~~~~~~~~~
	 \RC{2}{1} =
	 \frac{1}{2}\begin{pmatrix} 
		R_{\rm TM}^{(2,1)} - R_{\rm TE}^{(2,1)} & R_{\rm TM}^{(2,1)} + R_{\rm TE}^{(2,1)} \\
		R_{\rm TM}^{(2,1)} + R_{\rm TE}^{(2,1)} & R_{\rm TM}^{(2,1)} - R_{\rm TE}^{(2,1)} 
	\end{pmatrix},
	\label{R_dielectric_RL}
\end{align}
and the transmission matrix is given by
\begin{align}	
	\begin{pmatrix} \mathcal E_R' \\ \mathcal E_L'  \end{pmatrix}
	= \TC{2}{1}
	\begin{pmatrix} \mathcal E_R \\ \mathcal E_L \end{pmatrix},
	~~~~~~~~~~~~
	\TC{2}{1} =
    \begin{pmatrix} 
    T^{(2|1)}_{\rm TM} + T^{(2|1)}_{\rm TE} &
    T^{(2|1)}_{\rm TM} - T^{(2|1)}_{\rm TE} 
    \\ 
    T^{(2|1)}_{\rm TM} - T^{(2|1)}_{\rm TE} & 
    T^{(2|1)}_{\rm TM} + T^{(2|1)}_{\rm TE}
    \end{pmatrix}.
\end{align}

Finally we briefly comment on the case of perfect conductor.
Let us suppose that the material 2 is a perfect conductor.
Then the boundary condition  for $\vec E_\parallel$ (\ref{boundary_E}) and $\vec B_\perp$ (\ref{boundary_B}) with the assumption $\vec E_2 = \vec B_2 = 0$ easily leads to the reflection matrix
\begin{align}
	 \RC{2}{1} = \begin{pmatrix} 1 & 0 \\ 0 & 1  \end{pmatrix},
\end{align}
in the same basis as (\ref{R_dielectric_RL}).
A convenient way to connect this result to (\ref{R_dielectric_RL}) is to formally take $\epsilon_2 \to -\infty$ so that $k'= \sqrt{\epsilon_2 \omega} \sim i\infty$ and hence $R_{\rm TM}^{(2,1)} = 1$ and $R_{\rm TE}^{(2,1)} = -1$.

\subsection{Weyl semimetal}

Next let us consider the case of Weyl semimetal, in which there is a separation of the node in the dispersion relation, represented by the vector $\vec b$.
Since the Weyl semimetal is non-reciprocal due to the vector $\vec b$, a care is needed to correctly derive the reflection and transmission matrix.

\subsubsection*{Polarization of electromagnetic waves in the Weyl semimetal}

We first summarize some basic properties of electromagnetic waves in the Weyl semimetal. For more details, see e.g. Refs.~\cite{Wilson:2015wsa,Farias:2020qqp,Ema:2023kvw}.
The photon dispersion relation in the Weyl semimetal is given by
\begin{align}
	\omega^2 &= \vec k^2 + \frac{b^2}{2} \pm \frac{b}{2}\sqrt{b^2 + 4k_z^2}
	\nonumber \\
	 &=k_\parallel^2 + \left( \sqrt{k_z^2+ \frac{b^2}{4}} \pm\frac{b}{2}\right)^2.
	 \label{omega2_weyl}
\end{align}
For given $\omega$, there are two solutions for $k_z$:
\begin{align}
	(k_z^{\pm} )^2 = \kappa_z (\kappa_z\pm b).
\end{align}
where $\kappa_z \equiv \sqrt{\omega^2- k_x^2}$. 
The electric field $\vec E = \vec{\mathcal E} e^{i(\vec k\cdot\vec x-\omega t)}$ in the Weyl semimetal satisfies
\begin{align}
	\mathcal M
	\begin{pmatrix} 
		\mathcal E_x \\ \mathcal E_y \\ \mathcal E_z
	\end{pmatrix} = 0,~~~~~~
	\mathcal M \equiv
	\begin{pmatrix} 
		-\omega^2 + k^2-k_x^2 & i\omega b & -k_xk_z \\
		- i\omega b & -\omega^2 + k^2 &  0 \\
		-k_xk_z & 0 & -\omega^2 + k^2-k_z^2
	\end{pmatrix}.
\end{align}
By solving this, we obtain the polarization vector for the $\pm$ mode as
\begin{align}
	\hat e_{\pm} = \frac{1}{\sqrt 2 N_\pm}\left( \kappa_z^2 \hat x \pm i\omega \kappa_z \hat y-k_x k_z^\pm \hat z\right),
\end{align}
where $N_\pm$ is the normalization constant to make $\hat e_\pm$ the unit vector.  
Similarly, polarization vector for the left-moving mode $\vec k'' = (k_x,0,-k_z^\pm)$ is
\begin{align}
	\hat e''_{\pm} = \frac{1}{\sqrt 2 N_\pm}\left( -\kappa_z^2 \hat x \mp i\omega \kappa_z \hat y-k_x k_z^\pm \hat z\right).
\end{align}
Note that $\hat e_{+}$ and $\hat e''_{-}$ are (roughly) right-circular polarized waves and $\hat e_{-}$ and $\hat e''_{+}$ are left-circular polarized waves. 
It may be also convenient to define the basis vector for the magnetic field. 
For the right-moving mode, the polarization basis is given by
\begin{align}
	\hat e^{(B)}_{\pm} = \hat k_{\pm}\times \hat e_{\pm} =  \frac{\omega}{\sqrt 2 N_\pm k_\pm}\left( \mp i\kappa_z k_z^\pm \hat x + \omega k_z^\pm \hat y \pm ik_x \kappa_z \hat z\right).
\end{align}
The polarization vector for the left-moving mode $\vec k'' = (k_x,0,-k_z^\pm)$ is
\begin{align}
	\hat e^{''(B)}_{\pm} = \hat k''_{\pm}\times \hat e_{\pm} =  \frac{\omega}{\sqrt 2 N_\pm k_\pm}\left( \mp i\kappa_z k_z^\pm \hat x + \omega k_z^\pm \hat y \mp ik_x \kappa_z \hat z\right).
\end{align}

\begin{figure}
\begin{center}
    \includegraphics[width=15cm]{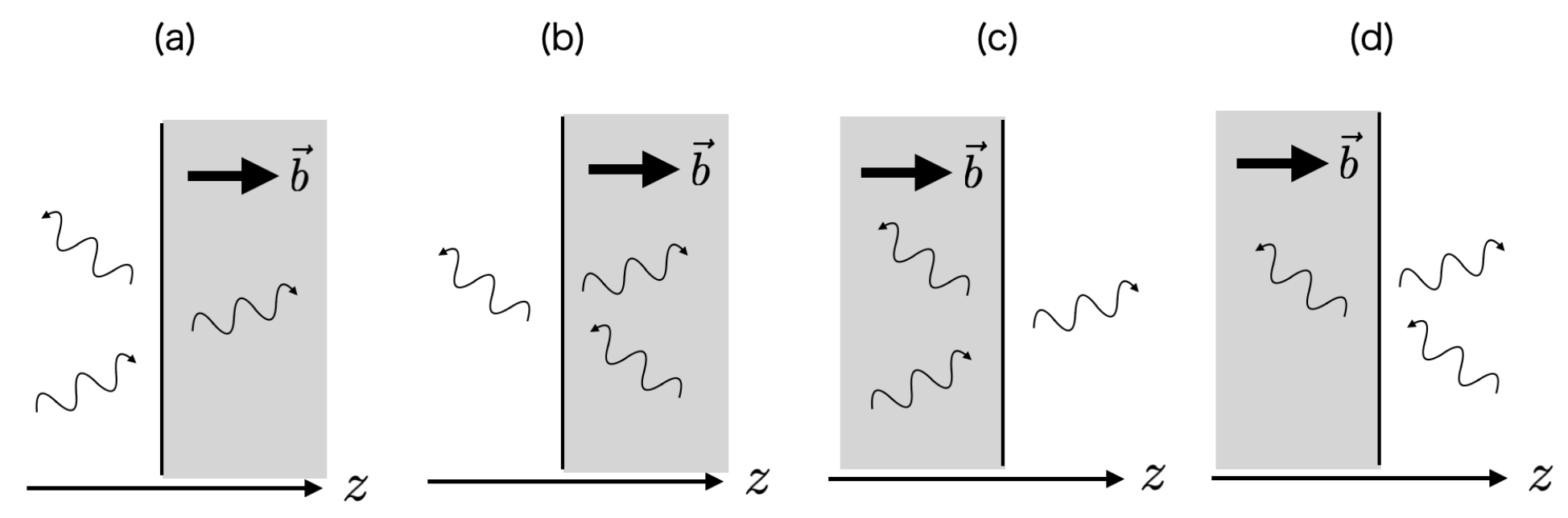}
    \end{center}
    \caption{Schematic pictures for reflection and transmission waves for Weyl semimetal.}
    \label{fig:pol}
\end{figure}

\subsubsection*{Reflection and transmission coefficients}

In the following we derive reflection/transmission coefficients for the case (a)-(d) shown in Fig.~\ref{fig:pol}.

\paragraph{Case (a)}
We consider the case shown in (a) of Fig.~\ref{fig:pol}.
The wave vector in the vacuum (Weyl semimetal) regions is denoted by $\vec q$ ($\vec k_\pm$). 
The incoming, reflected, and transmitted waves take the form of
\begin{align}
	&\vec E_i(\vec x) =e^{i(\vec q\cdot \vec x - \omega t)}  \left( \mathcal E_{R} \hat e_{R} +  \mathcal E_{L} \hat e_{L}   \right) ,\\
	&\vec E_r(\vec x) =e^{i(\vec {q''}\cdot \vec x - \omega t)} \left( \mathcal E''_{R} \hat e''_{R} + \mathcal E''_{L} \hat e''_{L}  \right),\\
	&\vec E_t(\vec x) =e^{i(\vec k_+\cdot \vec x - \omega t)}\mathcal E_{+} \hat e_{+} 
	+ e^{i(\vec k_-\cdot \vec x - \omega t)}\mathcal E_{-} \hat e_{-}.
\end{align}
Magnetic fields are obtained by $\vec B_i= \frac{\vec q\times \vec E_i}{\omega}$ and so on.
As noted in Sec.~\ref{app:dielectric}, we should have $\vec q_\parallel = \vec q''_\parallel = \vec k_{+\parallel} = \vec k_{-\parallel}$ and $q''_z = -q_z$ from the continuity at the boundary.
By solving boundary conditions, we obtain the reflection/transmission matrix as 
\begin{align}	
	\begin{pmatrix} \mathcal E_L'' \\ \mathcal E_R''  \end{pmatrix}
	= \RC{2}{1}
	\begin{pmatrix} \mathcal E_R \\ \mathcal E_L \end{pmatrix},
	~~~~~~~~~~~~
	 \RC{2}{1} =
	 \begin{pmatrix} 
		R_+ & 0 \\
		0 & R_-
	\end{pmatrix},
\end{align}
and
\begin{align}	
	\begin{pmatrix} \mathcal E_+ \\ \mathcal E_-  \end{pmatrix}
	= \TC{2}{1}
	\begin{pmatrix} \mathcal E_R \\ \mathcal E_L \end{pmatrix},
	~~~~~~~~~~~~
	 \TC{2}{1} =
	 \begin{pmatrix} 
		T_+ & 0 \\
		0 & T_-
	\end{pmatrix}.
\end{align}
where $R_\pm$ and $T_\pm$ are given in Eq.~(\ref{eq:RT_Weyl}).

\paragraph{Case (b)}
Next we consider the opposite propagation direction: the case of incoming photon from Weyl semimetal, which is reflected by the vacuum as schematically shown in (b) of Fig.~\ref{fig:pol}.
\begin{align}
	&\vec E_i(\vec x) =e^{i(\vec k''_+\cdot \vec x - \omega t)}\mathcal E''_{+}\hat e''_{+} + e^{i(\vec k''_-\cdot \vec x - \omega t)}\mathcal E''_{-}\hat e''_{-},\\
	&\vec E_r(\vec x) =e^{i(\vec k_+\cdot \vec x - \omega t)}\mathcal E_{+} \hat e_{+}+ e^{i(\vec k_-\cdot \vec x - \omega t)}\mathcal E_{-} \hat e_{-},\\
	&\vec E_t(\vec x) =e^{i(\vec {q''} \cdot \vec x - \omega t)} \left (\mathcal E''_{R} \hat e''_{R} + \mathcal E''_{L} \hat e''_{L}\right).
\end{align}
By solving the boundary conditions, we obtain the reflection/transmission matrix as 
\begin{align}	
	\begin{pmatrix} \mathcal E_+ \\ \mathcal E_-  \end{pmatrix}
	= \RC{1}{2}
	\begin{pmatrix} \mathcal E_+'' \\ \mathcal E_-'' \end{pmatrix},
	~~~~~~~~~~~~
	 \RC{1}{2} =
	 \begin{pmatrix} 
		-R_+ & 0 \\
		0 & -R_-
	\end{pmatrix},
\end{align}
and
\begin{align}	
	\begin{pmatrix} \mathcal E_L'' \\ \mathcal E_R''  \end{pmatrix}
	= \TC{1}{2}
	\begin{pmatrix} \mathcal E_+'' \\ \mathcal E_-'' \end{pmatrix},
	~~~~~~~~~~~~
	 \TC{1}{2} =
	 \begin{pmatrix} 
		T_+' & 0 \\
		0 & T_-'
	\end{pmatrix},
\end{align}
where $R_\pm$ and $T'_\pm$ are given in Eq.~(\ref{eq:RT_Weyl}).
Combined with the result of the case (a), we can check that the relation (\ref{eq:rel_jp1toj}) is satisfied:
$\left(\RC{1}{2}\right)^2 + \TC{2}{1}\TC{1}{2} = \mathbb{I} $ and $ \TC{2}{1} \RC{2}{1}+ \RC{1}{2} \TC{2}{1}=0$.

\paragraph{Case (c)}
Next we consider the case shown in (c) of Fig.~\ref{fig:pol}:
\begin{align}
	&\vec E_i(\vec x) =e^{i(\vec k_+\cdot \vec x - \omega t)}\mathcal E_{+}\hat e_{+} + e^{i(\vec k_-\cdot \vec x - \omega t)}\mathcal E_{-}\hat e_{-},\\
	&\vec E_r(\vec x) =e^{i(\vec k''_+\cdot \vec x - \omega t)}\mathcal E''_{+} \hat e''_{+}+ e^{i(\vec k''_-\cdot \vec x - \omega t)}\mathcal E''_{-} \hat e''_{-},\\
	&\vec E_t(\vec x) =e^{i(\vec q \cdot \vec x - \omega t)} \left (\mathcal E_{R} \hat e_{R} + \mathcal E_{L} \hat e_{L}\right).
\end{align}
By solving the boundary conditions, we obtain the reflection/transmission matrix as 
\begin{align}	
	\begin{pmatrix} \mathcal E_+'' \\ \mathcal E_-''  \end{pmatrix}
	= \RC{2}{1}
	\begin{pmatrix} \mathcal E_+ \\ \mathcal E_- \end{pmatrix},
	~~~~~~~~~~~~
	 \RC{2}{1} =
	 \begin{pmatrix} 
		-R_+ & 0 \\
		0 & -R_-
	\end{pmatrix},
\end{align}
and
\begin{align}	
	\begin{pmatrix} \mathcal E_R\\ \mathcal E_L  \end{pmatrix}
	= \TC{2}{1}
	\begin{pmatrix} \mathcal E_+ \\ \mathcal E_- \end{pmatrix},
	~~~~~~~~~~~~
	 \TC{2}{1} =
	 \begin{pmatrix} 
		T_+' & 0 \\
		0 & T_-'
	\end{pmatrix}.
\end{align}

\paragraph{Case (d)}
Finally we consider the case shown in (d) of Fig.~\ref{fig:pol}:
\begin{align}
	&\vec E_i(\vec x) =e^{i(\vec q''\cdot \vec x - \omega t)}\left(\mathcal E''_{L}\hat e''_{L} + \mathcal E''_{R}\hat e''_{R}\right),\\
	&\vec E_r(\vec x) =e^{i(\vec q\cdot \vec x - \omega t)}\left(\mathcal E_{R} \hat e_{R} + \mathcal E_{L} \hat e_{L}\right),\\
	&\vec E_t(\vec x) =e^{i(\vec {k_+''} \cdot \vec x - \omega t)} \mathcal E''_{+} \hat e''_{+} + e^{i(\vec {k_-''} \cdot \vec x - \omega t)} \mathcal E''_{-} \hat e''_{-} .
\end{align}
By solving the boundary conditions, we obtain the reflection/transmission matrix as 
\begin{align}	
	\begin{pmatrix} \mathcal E_R \\ \mathcal E_L  \end{pmatrix}
	= \RC{1}{2}
	\begin{pmatrix} \mathcal E_L'' \\ \mathcal E_R'' \end{pmatrix},
	~~~~~~~~~~~~
	 \RC{1}{2} =
	 \begin{pmatrix} 
		R_+ & 0 \\
		0 & R_-
	\end{pmatrix},
\end{align}
and
\begin{align}	
	\begin{pmatrix} \mathcal E_+'' \\ \mathcal E_-''  \end{pmatrix}
	= \TC{1}{2}
	\begin{pmatrix} \mathcal E_L'' \\ \mathcal E_R'' \end{pmatrix},
	~~~~~~~~~~~~
	 \TC{1}{2} =
	 \begin{pmatrix} 
		T_+ & 0 \\
		0 & T_-
	\end{pmatrix}.
\end{align}
Again, combined with the result of the case (c), we can check that the relation (\ref{eq:rel_jp1toj}) is satisfied:
$\left(\RC{1}{2}\right)^2 + \TC{2}{1}\TC{1}{2} = \mathbb{I}$ and $ \TC{2}{1} \RC{2}{1}+ \RC{1}{2} \TC{2}{1}=0$.

Note that the result of (d) is obtained from that of (a) by the following replacement: $L\leftrightarrow R$ and $+\leftrightarrow -$ (which follows from $\vec b \leftrightarrow -\vec b$). The same is true also for the case (b) and (c).

\bibliographystyle{utphys}
\bibliography{ref}

\end{document}